\documentclass[10pt,twocolumn,letterpaper]{article}

\usepackage[pagenumbers]{cvpr} 

\usepackage{graphicx}
\usepackage{amsmath}
\usepackage{amssymb}
\usepackage{booktabs}
\usepackage{animate}
\usepackage{enumitem}

\usepackage{hyperref}
\hypersetup{pagebackref,breaklinks,colorlinks}

\usepackage[capitalize]{cleveref}
\crefname{section}{Sec.}{Secs.}
\Crefname{section}{Section}{Sections}
\Crefname{table}{Table}{Tables}
\crefname{table}{Tab.}{Tabs.}

\def\cvprPaperID{7426} 
\def\confName{CVPR}
\def\confYear{2022}

\newcommand{\op}[1]{{\color{black}#1}}
\newcommand{\opb}[1]{{\color{black}#1}}
\begin{document}

\title{Video Reconstruction from a Single Motion Blurred Image \\ using Learned Dynamic Phase Coding}

\author{Erez Yosef\\
Tel Aviv University, Israel\\
{\tt\small erez.yo@gmail.com}
\and
Shay Elmalem\\
Tel Aviv University, Israel\\
{\tt\small shay.elmalem@gmail.com}
\and
Raja Giryes\\
Tel Aviv University, Israel\\
{\tt\small raja@tauex.tau.ac.il}
}
\maketitle


\begin{abstract}

Video reconstruction from a single motion-blurred image is a challenging problem, which can enhance the capabilities of existing cameras. Recently, several works addressed this task using conventional imaging and deep learning. Yet, such purely-digital methods are inherently limited, due to direction ambiguity and noise sensitivity. Some works proposed to address these limitations using non-conventional image sensors, however, such sensors are extremely rare and expensive. To circumvent these limitations with simpler means, we propose a hybrid optical-digital method for video reconstruction that requires only simple modifications to existing optical systems. We use a learned dynamic phase-coding in the lens aperture during the image acquisition to encode the motion trajectories, which serve as prior information for the video reconstruction process. The proposed computational camera generates a sharp frame burst of the scene at various frame rates from a single coded motion-blurred image, using an image-to-video convolutional neural network. We present advantages and improved performance compared to existing methods, using both simulations and a real-world camera prototype. \opb{We extend our optical coding also to video frame interpolation and present robust and improved results for noisy videos.} \looseness=-1

\end{abstract}


\maketitle

\newcommand\offsetx{.93}
\newcommand\offsety{1.74}

\section{Introduction}
\label{sec:intro}

Modern cameras are required to satisfy two conflicting requirements: to provide excellent imaging performance while decreasing the space and weight of the system. To address this inherent contradiction, novel design methods attempt to harness fundamental imaging limitations and leverage them as a design advantage. One such example is motion blur, which is a known limitation in photography of dynamic scenes. It is caused due to objects' movements during exposure, whose duration is set according to the lighting conditions and noise requirements. As most scenes are dynamic, light from moving objects is accumulated by the sensor in several consecutive pixels along their trajectory, resulting in image blur. 
Although blur is an undesirable effect, in this work, we use it for video generation from a single image.

In contrast to motion deblurring methods that aim at sharp image reconstruction, in video generation the goal is to exploit this 'artifact' for reconstructing a sharp video frame burst that represents the scene at different times during acquisition.
Yet, as signal averaging in the acquisition process eliminates the motion direction in the captured image, this task is highly ill-posed. The pioneering work of \cite{Jin_2018_CVPR} suggests a pairwise frames order invariant loss to mitigate this ambiguity. Yet, as the global motion direction is lost in the acquisition, the processing stage can only assume the direction of the motion for the video reconstruction but cannot really resolve the global direction ambiguity.

To overcome this deficiency, some works suggested capturing multiple frames with different exposures during the acquisition process \cite{Rengarajan2020PhotosequencingOM} or alternatively replacing the sensor with coded two-bucket \cite{Anupama,shedligeri2020unified} or event measurements \cite{Pan_2019_CVPR}. Yet, these solutions do not fit with a standard optical system or require capturing multiple images.

\begin{figure}[t]
\begin{subfigure}[b]{0.33\linewidth}
\centering
\includegraphics[width=1in ,bb=0 0 190 190]{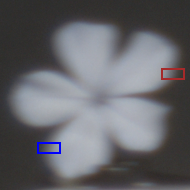}
   \caption{Coded-blurred image}
   \label{fig:teaser_blur}
\end{subfigure}
\captionsetup[subfigure]{width=80pt}
\begin{subfigure}[b]{0.33\linewidth}
\centering
\includegraphics[width=1in,bb=0 0 180 180]{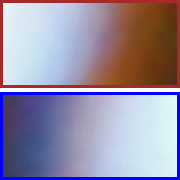}
    \vspace*{-6pt}%
   \caption{Motion-color cues}
   \label{fig:teaser_color_cues}

\end{subfigure}
    \begin{subfigure}[b]{0.31\linewidth}
    \centering
    \animategraphics[loop,width=1in,poster=12, bb=0 0 190 190]{8}{Figs/teaser/learned_rflr_2_s0_f}{0}{24}
    \vspace*{-18pt}
    \caption{Reconstructed video}
   \label{fig:teaser_vid} 
\end{subfigure}

    \caption{\textbf{Method demonstration.} (a) A flower moving left captured using our dynamic phase-coded camera, which embeds (b) color-motion cues in the intermediate image. These cues guide our image-to-video reconstruction CNN, resulting in a (c) sharp video of the scene (play the video by clicking on (c) in Adobe Reader).}
    \label{fig:teaser}
\end{figure}

 \begin{figure*}[t]
 \begin{center}
   \includegraphics[width=\linewidth]{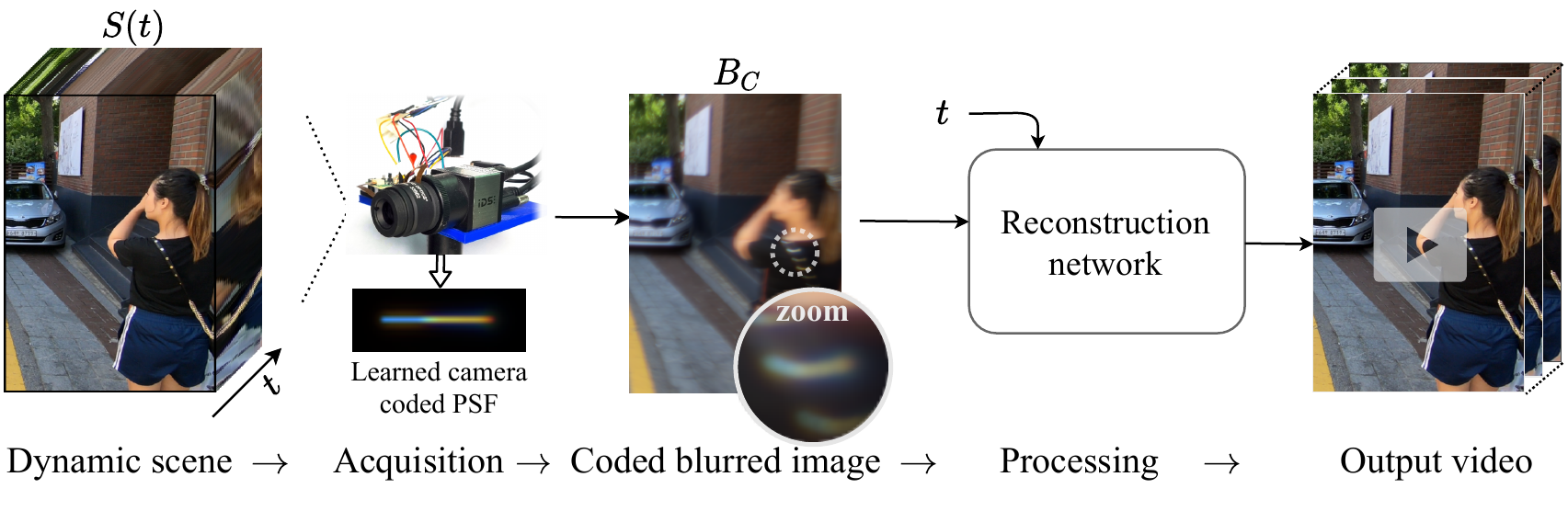}

   \caption{{\bf{Overview of our suggested method.}} An acquisition of a dynamic scene using our dynamically phase-coded camera provides an intermediate image $B_C$ which contains scene dynamics cues in its coded motion-blur. We reconstruct sharp video frames of the scene at desired timesteps $t$ from the single coded-blurred image $B_C$ using a time-dependent CNN. The optical coding parameters are jointly optimized with the reconstruction network weights using end-to-end learning.}
   \label{fig:system_diagram}
\end{center}
\end{figure*}

\noindent{\bf{Contribution.}} To overcome the limitations of conventional cameras in dynamic scenes acquisition, we suggest a computational coded-imaging approach (see \cref{fig:teaser,fig:system_diagram}) that can be easily integrated in many conventional cameras (equipped with a focusing mechanism) by just adding a phase-mask to their lens (which is a simple process). Joint operation of the phase-mask and focus variation during exposure generates a dynamic phase coding, which encodes scene motion information in the intermediate image as chromatic cues. The cues are generated by the PSF nature of our solution (plotted in \cref{fig:psf}), which encodes the beginning of the movement by blue and the end by red, e.g., see the zoomed left and right edges of the moving flower in \cref{fig:teaser_color_cues} (enhanced for visualization). These cues serve as guidance for generating a video of the scene motion by post-processing the captured coded image (see \cref{fig:teaser_vid}). 

Our method is capable of generating a sharp frame at any user-controlled time in the exposure interval. Therefore, a video burst at any user-desired frame rate can be produced from a single coded image.
The proposed coding and reconstruction approach is based on a learnable imaging layer and a Convolutional Neural Network (CNN), which are jointly end-to-end optimized; the learnable imaging layer simulates the physical image acquisition process by applying the coded spatiotemporal point spread function (PSF), and the CNN reconstructs the sharp frames from the coded image.

The main contributions of our method are:
\begin{itemize}[leftmargin=*]
\itemsep0em 
\item \op{A learning based coding method, which requires only a conventional sensor and a lens that has a focusing mechanism and is equipped with a simple add-on optical phase-mask. 
\item The learned code allows using low-power neural networks while achieving high quality video reconstruction from a single image.
\item A novel neural architecture that leads to a flexible and modular video from a single image reconstruction, which can be easily adjusted to any desired frame rate video by a simple change in the neural network parameters that do not require re-training.}
\item End-to-end optimization framework of optical and digital processing parameters for dynamic scene acquisition\op{, which accurately models a spatio-temporal acquisition process.} 
\item Improved video from motion reconstruction, with unambiguous directionality, higher accuracy and lower noise sensitivity, tested in both simulation and real-world experiments.
\op{\item A novel video perceptual loss for training neural networks and a metric for video reconstruction evaluation that takes into account both spatial and temporal information.}
\opb{\item A method for video frame interpolation using the proposed coding that achieves improved results for real-world dynamic scenes.
\item SNR tradeoff analysis for dynamic scenes acquisition and frame interpolation.
}
\end{itemize}

\section{Related Work}
\label{sec:prevWork}

Given a motion blurred image, various methods attempted to reconstruct a sharp image of the scene from it. Some techniques were developed for the case of conventional imaging and then the reconstruction is only computational, which recently is usually based on training a neural network for this task \cite{Zhang2018_Scene_Deblurring, tao2018srndeblur, Kupyn_2018_CVPR, Nah_2017_CVPR}. 
Other holistic design approaches utilize a computational imaging strategy to encode motion information in an intermediate image and recover the sharp image using corresponding post-processing methods \cite{Raskar2006CodedEP, Levin_motionInvariant,Levin_orthParab, CAVE_0040, Srinivasan_2017_CVPR, R2017_ICCV}.


\begin{table*}
  \centering
  \begin{tabular}{@{}lcccc@{}}
    \toprule
    Method & Acquisition & Input & Output size \\
    \midrule
    Computational only \cite{Jin_2018_CVPR,Purohit2019BringingAB,Zhang2020EveryMM} & 
    conventional & 
    image & 
    fixed \\
    Multiple exposures \cite{Rengarajan2020PhotosequencingOM}& 
    short-long-short exposures & 
    3 images & 
    dynamic \\
    Coded two-bucket \cite{Anupama,shedligeri2020unified}& 
    C2B sensor &
    two coded images & 
    fixed \\
    Event camera \cite{Pan_2019_CVPR} & 
    event sensor &  
    50-100 events vector &  
    dynamic\\
    Proposed method & 
    dynamic phase coding &
    coded image &  
    dynamic\\
    \bottomrule
  \end{tabular}
  \caption{Overview of existing solutions for video reconstruction from a motion-blurred scene.}
  \label{table:related_work}
  \label{tab:prevWork}
\end{table*}

The problem of video reconstruction from a single image takes motion deblurring a step forward by attempting to reconstruct a frame burst of the dynamic scene that resulted in the blurred image, and not only the central sharp frame (see \cref{tab:prevWork} for an overview). Some works are based on images taken using a conventional camera and apply processing-only methods to obtain a frame burst of the scene. However, without optical coding, this problem is highly ill-posed as even if the edges and textures are reconstructed perfectly, various motion permutations can generate the same motion blurred image (e.g., see Fig.~2 in \cite{Jin_2018_CVPR}). Thus, coded imaging approaches were proposed to acquire additional information about the scene dynamics and achieve higher quality results.  

\noindent{\textbf{Conventional imaging based methods.}} Generating a video sequence from a single motion blurred image is a challenging task: since the temporal order of the reconstructed frames is ambiguous, the problem is highly ill-posed.
\cite{Jin_2018_CVPR} address the temporal order ambiguity and present a pioneering approach for this task using several reconstruction networks and a novel pairwise frames order-invariant loss. Their suggested method consists of iterative generation of seven sequential frames of the scene, starting from the central frame reconstruction and proceeding to the edge frames of the dynamic scene using the preceding reconstruction results. The method's architecture limits the reconstruction to only seven frames of the scene in the exposure interval, and it uses three different trained models for the reconstruction process. \cite{Purohit2019BringingAB} present a solution for video reconstruction using motion representations of the scene learned by recurrent video autoencoder network. \cite{Zhang2020EveryMM} suggested a detail-aware network using a cascaded generator. \op{\cite{learning_to_deblur_and_rotate} suggest a solution for rendering sharp video of a face from new viewpoints from a single motion-blurred image.} All of these methods suffer from the inherent motion direction ambiguity, and their reconstruction performance is more sensitive to noise (as discussed in \cite{Cossairt_compImagAnlz,OSA_compImgRev} and empirically presented in \cref{sec:exp}).

\noindent{\textbf{Coded imaging based methods.}} To handle the inherent limitations of conventional imaging, some works adopted computational photography methods for image deblurring and video frames recovery. \cite{Raskar2006CodedEP} introduced an amplitude coded exposure technique using fluttered shutter for motion deblurring. This method performs a temporal binary amplitude coding, resulting in a wider frequency response, which is utilized for improved motion deblurring results. \cite{Levin_motionInvariant,Levin_orthParab} presented a parabolic motion camera with motion invariant PSF utilized for non-blind motion deblurring. Both of these approaches are limited to the reconstruction of a single image. \op{\cite{5585094} presented a spatial-temporal coding exploiting the rolling shutter of CMOS  sensor.}
Dynamic phase coding in the lens aperture for motion coding was presented by \cite{Shay2020phase} for motion deblurring. This coding embeds motion cues in the intermediate image for improved deblurring performance.
For video restoration from a single coded-blurred image, several approaches had been presented, such as using an event camera \cite{Pan_2019_CVPR}, or coded two-bucket (C2B) sensor \cite{Anupama, shedligeri2020unified, C2B}, which both require a non-conventional sensor or lensless imaging and rolling shutter effect \cite{Antipa2019VideoFS}, which omits the lens and therefore changes the entire imaging concept even for static scenes. We adopt the coding method of \cite{Shay2020phase}, which is based on a commercial sensor and lens (with focusing mechanism), equipped with a simple add-on optical element, which allows unambiguous motion cues encoding. Different from \cite{Shay2020phase} that performs only image deblurring, we aim at reconstructing a video of the motion in the scene from a single motion blurred image, which is a vaster and more ill-posed task. 

A closely related problem is a reconstruction of a sharp and high frame rate video from motion blurred and low frame rate video, using either processing of conventional camera videos \cite{Jin_2019_CVPR, Rengarajan2020PhotosequencingOM, NEURIPS2020_UTI, Shen_2020BIN_CVPR} or computational imaging methods \cite{FlutterShutterVideo, 6552198, Llull_Coded_aperture, Event-Driven-Video}. These methods require a video input (which enables solving the direction ambiguity) and are not applicable for single image input.

\noindent{\textbf{Deep optics.}} As the end-to-end backpropagation-based optimization process of deep models proved itself to be very efficient for various tasks, its power was also harnessed for optical design, either for a standalone optical system design process or jointly with a post-processing algorithm (for recent review on this topics see \cite{compImagDL_revOsa,OpticsDL_revNature}). Specifically for enhanced optical imaging applications, this scheme had been presented for extended depth of field \cite{EDOF_DL,Gordon_EDOF,Ugur_EDOF, Tan2021CodedStereoLP}, depth estimation \cite{Depth_2018,Yicheng_Depth,Gordon_Depth,Chen2020AutoTuningSL}, high dynamic range \cite{Gordon_HDR,Heide_HDR}, ray tracing \cite{raytrace1, raytrace2}, other tasks\cite{SRSPAD, Sheinin:2021:Deconv, sssszxvzxvz, Dowski1999WavefrontCA, Robinson:06} and several microscopy applications \cite{Waller_Illum,Waller_MiniscopeS,Shechtmann_MultiChan,Shechtman_rev}, to name a few. Yet, it was not considered for the problem of video from blur.
\section{Method}
\label{sec:camDesign}

As our goal is to reconstruct video frames from a motion blurred image of the scene, we engineer the camera's PSF to encode cues in the motion blur of dynamic objects. The coded PSF is achieved using a spatiotemporal dynamic phase coding in the lens aperture, which results in motion-coded blur. The coded blur serves as prior information for the image-to-frames CNN, trained to generate sharp video frames from the coded image. Utilizing the end-to-end optimization ability, the optical coding process is modeled as a layer in the model, and its physical parameters are optimized along with the conventional CNN layers in a supervised manner.
The learned optical coding is then implemented in a prototype camera, and images taken using it are processed using the digital processing layers of the CNN.

\subsection{Camera Dynamic Phase Coding} \label{method_camera_dpc}
Moving objects in a scene during exposure result in motion blur, as the light from a moving object is integrated in different pixels along the motion trajectory. In addition, both static and dynamic objects are blurred by the lens PSF which is never perfect (due to aberrations/diffraction etc.). 
This imaging process is formulated in \cref{camera_eq}; the two-dimensional PSF is spatially convolved with the instantaneous scene at any $dt$ and integrated during exposure, $B$ is the acquired blurred image\footnote{All images mentioned are in the linear regime (signal space), i.e. before any non-linear transformations such as gamma correction.}, $T$ is the exposure time, $S(t)$ and $h$ denote the instantaneous sharp scene and the PSF respectively, and $(\underset{Sp}{*})$ denotes the spatial convolution operator (the spatial coordinates are omitted for ease of notation). 
\begin{equation}\label{camera_eq}
B = \frac{1}{T}\int_{0}^{T} \left(h \underset{Sp}{*} S(t)\right)\,dt \simeq \frac{1}{N}\sum_{n=1}^{N} \left(h \underset{Sp}{*}S(n)\right)
\end{equation}
The averaging nature of image sensors results in the loss of the motion direction, which introduces inherent ambiguity. 
Also, as every object moves independently of others, general motion blur is shift-variant. Thus, video reconstruction from undirected motion blur is a highly ill-posed task.

To address both issues, we implement a coded lens designed to embed motion cues in the acquired image, and the prior knowledge about the camera's time-variant behavior serves as guidance to the reconstruction process of the video burst.
We adopt dynamic phase coding in the lens aperture, similar to the motion deblurring method presented by \cite{Shay2020phase}. This method is based on a spatiotemporally coded PSF that encodes motion information in the intermediate image without attenuation of the signal compared to amplitude coding methods such as \cite{Raskar2006CodedEP,FlutterShutterVideo}.\op{ Such an approach improves the signal to noise ratio (SNR) compare to amplitude coding methods.} Since video reconstruction from a blurred image is a more ill-posed task than deblurring, we improve the imaging method by optimizing the coding parameters for our task using end-to-end learning of them with the reconstruction network.

The camera PSF is generated using a conventional camera equipped with a simple add-on phase-mask; the temporal coding is achieved using a joint operation of the static phase-mask designed to introduce color-focus cues, and a dynamic focus sweep performed during exposure (using a simple focusing mechanism). The phase-mask (originally designed for depth estimation \cite{Depth_2018} and extended depth of field imaging \cite{EDOF_DL}) introduces a predesigned chromatic aberration to the lens, generating a controlled dependence between the defocus condition and the color distribution of the PSF. \op{Based on Fourier optics, the PSF of the camera is computed conditioned on the mask specifications, the wavelength and the defocus condition as previous works (extended in \cref{sec:PSFcomputation}).} 
To get a time-varying PSF the defocus condition (denoted as $\psi$) is changed during exposure, and a temporally coded PSF (denoted as $h(\psi(t))$) is achieved. The instantaneous scene $S(t)$ is spatially convolved with the corresponding PSF $h(\psi(t))$, resulting in the motion-coded image $B_c$ described in the following formula:
\begin{eqnarray}
\label{coded_camera_eq}
\nonumber && \hspace{-0.9in}
B_c = \frac{1}{T}\int_{0}^{T} \left(h(\psi(t)) \underset{Sp}{*} S(t)\right)\,dt \simeq \\ && \frac{1}{N}\sum_{n=1}^{N} \left(h(\psi(n)) \underset{Sp}{*}S(n)\right).
\end{eqnarray}

Using the proposed spatiotemporally coded imaging scheme, the dynamics of the scene are encoded in the intermediate image acquired by the camera. Moving objects are smeared in the image with color cues along their trajectories, based on the spatiotemporal PSF $h(\psi(t))$. The acquired coded image is then fed to the reconstruction network trained to decode these cues as guidance for improved video reconstruction. \cref{fig:system_diagram} presents these steps visually.

\subsection{PSF Design} \label{psf_design}
\op{The time varying PSF conditioned on the defocus parameter $\psi$ is computed based on Fourier optics and described in \cref{sec:PSFcomputation}.}
To achieve optimal motion cues encoding in the intermediate image, the imaging process ( \cref{sec:PSFcomputation}) is modeled as a learnable layer (with corresponding forward and backward models using automatic differentiation). The dynamic phase coding acquisition is simulated using the phase-mask characteristics and the focus variation parameters. The defocus parameters are optimized in the end-to-end training process along with the CNN layers, while the phase mask design is fixed (constant) in our setup and described in \cref{sec:Opticalsystem}. The initialization method of the optical parameters was tested using different approaches in the acceptable physical range, including linear as \cite{Shay2020phase}, random, and even some approaches combined with periodic functions (i.e. sine). The linear initialization produced the best convergence results over all other attempts. Note that as we are using temporal phase coding, we do not change the intensity during the change of the focus and we optimize only the focus as a function of time.

 \begin{figure}[t]
  \begin{center}
   \includegraphics[trim=0.2cm 0cm 0.6cm 0cm , width=\linewidth]{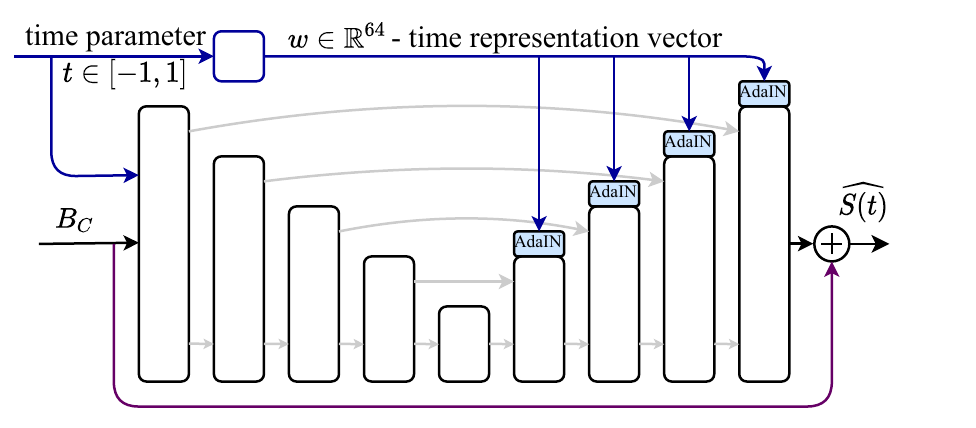}

   \caption{{\bf{Network architecture.}} Our CNN is based on the UNet \cite{Unet2015} model, with the coded blurred image and a time parameter as inputs and the sharp reconstructed frame at the output (see \cref{our_network_eq}). The decoder part is controlled by the time parameter (using AdaIN \cite{Huang_2017_ICCV}), to set the relative time of the reconstructed frame.}
   \label{fig:unet_diagram}
   \end{center}
\end{figure}

\subsection{Reconstruction Network} \label{sec:cnnArch}

Our proposed model for video frames reconstruction from a coded motion-blurred image is based on a single time-dependent convolutional neural network (CNN) with AdaIN mechanism \cite{Huang_2017_ICCV}. Our model inputs are the coded-blurred intermediate image and a normalized time parameter $t\in[-1,1]$. The time parameter controls the relative time of the generated sharp frame in the normalized exposure time interval. The output of the model is the estimated sharp scene frame at time $t$, denoted as $\widehat{S(t)}$. Hence, the architecture is designed to reconstruct the scene at any desired instant in the exposure time interval, and thus to create a video at any desired frame rate. We denote it by
\begin{equation}\label{our_network_eq}
\widehat{S(t)} = f(B_c,\:t)\quad t\in[-1,1].
\end{equation}

Our reconstruction CNN (presented in \cref{fig:unet_diagram}) is based on the UNet architecture \cite{Unet2015}, consisting of a four levels encoder-decoder network structure with skip connections between the encoder to the decoder in each level. The double convolution blocks (presented in the original UNet architecture) are improved by adding skip-connections modifying them to the form of dense blocks \cite{Huang_2017_CVPR}. The output of the last layer of the model is added to the input image, such that the network learns only the residual correction required to reconstruct the desired frame.

The time parameter is used to reconstruct the frame corresponding to the desired normalized time in the exposure interval. It controls the network by both the AdaIN mechanism \cite{Huang_2017_ICCV} and concatenating it to the input as an additional channel \op{(by expanding the scalar value to image dimensions)}. 
To bridge between the shift-invariant convolutional operations of the CNN and the shift-variant (and scene dependent) motion blur that exists in our target application, we leverage positional encoding to add image position dependency to the model. We provide additional details on the architecture and these changes in the following.

\noindent{\textbf{Positional encoding.}} Assuming a general scene in which every object might move in a different direction and velocity, an intermediate image captured using our proposed coded lens will contain a shift-variant blur kernel, which is a composition of the color-temporal PSF coding and the spatial movement of the objects. Since convolutions are shift-invariant, we want to add a position dependency to the model, such that we can utilize the local information of the coding in the surrounding area that relates to the same object with the same motion characteristics and blurring profile. We adopt Fourier features to get a better representation of the position coordinates \cite{tancik2020fourfeat}. Similar to \cite{Metzer2021Z2PIR}, we add a positional dependency to the model by concatenating the Fourier features of the pixel coordinates as additional channels to the input.
Five log-linear spaced frequencies $\{w_j\}_{j=1}^{5}$ were sampled in the range [1,20] to generate 20 positional features in total for each pixel coordinate $(u,v)$ of the image. Each frequency $w_j$ contributes the following four positional features to each pixel using the normalized pixel coordinate in the range of [0,1]:
\begin{eqnarray}\label{posenc_eq}
\nonumber && \hspace{-0.4in} pos_j(u,v) =  \\&&  \hspace{-0.4in} [cos(w_j\cdot u), sin(w_j\cdot u),  cos(w_j\cdot v), sin(w_j\cdot v)].
\end{eqnarray}

\noindent{\textbf{Time encoding.}} To achieve a time-dependent CNN, the batch normalization layers in the UNet architecture are replaced with AdaIN layers \cite{Huang_2017_ICCV} controlled by a normalized time parameter. The exposure time interval is normalized to the range of $[-1,1]$ such that $t=0$ corresponds to the middle of exposure time. The time parameter $t\in[-1,1]$ is mapped to a higher dimension vector $w\in\mathbb{R}^{64}$, using an MLP network consisting of two sequential blocks of a linear layer followed by a leaky-ReLU activation function. The encoded time-representation vector $w$ is shared across all AdaIN layers and controls the mean and standard deviation of the features in each AdaIN layer.

In each AdaIN layer with an input $x$ of $p$ feature channels, the mean $\beta\in\mathbb{R}^{p}$ and standard deviation $\gamma\in\mathbb{R}^{p}$ are obtained from $w$ by designated MLP mapping networks with two layers of the same structure mentioned above. The AdaIN transformation (\cref{adain_norm}) is performed along the features dimension, where $\mu(x)$ and $\sigma(x)$ are computed across spatial dimensions (instance normalization).
\begin{equation}\label{adain_norm}
\hat{x} = \gamma\cdot\frac{x-\mu(x)}{\sigma(x)} + \beta.
\end{equation}

As our scheme is designed to utilize the optically encoded motion cues to generate a sharp frame in a relative time $t$, the encoder part of the UNet is generic, and we apply the temporally controlled AdaIN only on the decoder part of the architecture (as in \cref{fig:unet_diagram}). We set the encoder part of the UNet model to be time independent by performing instance normalization followed by a learnable affine transformation instead of the AdaIn blocks. In this setting, the encoder is optimized to encode more general information about the image and scene dynamics regardless of the normalized time parameter. The generic encoder and time-specific decoder design enable the network to converge better. Note though that we concatenate the time parameter to the input channels which contain the input image and the positional encoding features. This improves reconstruction performance as shown in the ablation in \cref{sec:ablation}.

\noindent{\textbf{Dataset.}} To train our network and evaluate its performance quantitatively, we used the REDS dataset \cite{Nah_2019_CVPR_Workshops}, consisting of scenes captured at 120 frames per second (FPS). To achieve smoother motion-blur simulation we used $\times 8$ frame interpolation using the DAIN method \cite{DAIN} (similarly to the process presented in \cite{Nah_2019_CVPR_Workshops}), to achieve video frames at 960 FPS. \opb{Inverse camera response function (CRF) was applied on the frames to convert them from gamma space to signal space, using the inverse crf transform given with the dataset.} To simulate the acquisition of a dynamic scene by our coded camera, the spatiotemporal PSF had been applied to 49 consecutive frames \opb{in signal space}, which were then averaged along the time axis as in Eq.~\eqref{coded_camera_eq} (where $N=49$). For performance comparison with \cite{Jin_2018_CVPR}, conventional camera images were simulated by only averaging the frames without applying the PSF.
Due to the applied frame interpolation, not all the 49 frames are true images; therefore only the seven real frames (in indices $n=8k,\; k \in [0,6]$) are used as our GT images for the training/validation/test metrics. 
For improved generalization, we add additive white Gaussian noise (AWGN) to the simulated blurred images in the signal space, which partially simulates the imaging process noise and improves the robustness of our model and generalization to the camera prototype (different noise levels were set according to the application, as discussed in \cref{sec:exp}).

\noindent{\textbf{Loss Functions.}} We use a linear combination of three losses for the training: pixel-values smooth-L1 loss ($l_{L1}$), perceptual loss ($l_{percep}$) using VGG features \cite{perceptual2016}, and a video-consistency perceptual loss ($l_{vid}$). Thus, our loss is
\begin{equation}\label{loss_eq}
l = \alpha_{L1}\cdot l_{L1}+\alpha_{percep}\cdot l_{percep}+\alpha_{vid}\cdot l_{vid}.
\end{equation}

The perceptual loss is a known practice for image reconstruction tasks \cite{perceptual2016}. In this loss, we compute the smooth-L1 distance between the VGG \cite{Simonyan2015VeryDC} features of the reconstructed image and the ground truth image.

\op{To improve temporal consistency and perceptuality between consecutive reconstructed video frames, we developed a video loss using a 3D convolution network over the video time-space volume. We use 3D-ResNet \cite{Tran2018ACL}, a spatiotemporal convolution network for video action recognition, and compare the network-extracted feature maps between the reconstruction and the GT videos. We used the output of the first three convolution layers of the 3D-ResNet network and averaged the smooth-L1 loss between the features of the ground truth video and the reconstructed video. More details are provided in \cref{sec:Vidloss_ap}.}

\section{Experiments}
\label{sec:exp}

As an experimental validation to our proposed approach, we first train our system (optical coding layer and reconstruction network) and evaluate our results quantitatively (while the optical coding process is simulated), and compare the performance to the previous work by \cite{Jin_2018_CVPR}. Following the satisfying simulative experiment, we built a prototype camera implementing our spatiotemporal coding and examined our method qualitatively (as pixelwise GT sharp frame bursts are almost impossible to acquire). Lastly, we present an ablation study for our architecture and used methods. Some of the results are presented below, and additional results are presented in Appendix~\ref{sec:Results} and the video.

\noindent{\textbf{Training details.}} We train our model on a training set consisting of 9,680 scenes for 40 epochs, with a batch size of 72 samples of patches in size 128x128x3 each. We used Adam optimizer \cite{Kingma2015AdamAM} with learning rate of $10^{-3}$ and weight decay of $10^{-8}$. The loss weightings (as defined as \cref{loss_eq}) are $\alpha_{L1}=1;\;\alpha_{percep}=0.1;\;\alpha_{vid}=0.1$. Additional 2460 scenes are dedicated for validation/testing, such that the quantitative reconstruction performance (\cref{sec:simRes}) was evaluated using 1,968 scenes dedicated for testing. In the optical coding layer we define a learnable defocus condition vector $\overline{\psi}\in\mathbb{R}^{49}$. \op{We optimize these focus sweep parameters of the camera, which defines the camera time-varying PSF following the computation presented in \cref{sec:PSFcomputation} , and the result obtained following the imaging Eq.~\eqref{coded_camera_eq}.} These parameters initialized linearly as discussed in \cref{psf_design}. To improve robustness we apply flip augmentations and add AWGN to the input image (1\% as in \cite{Jin_2018_CVPR} for \cref{sec:simRes}, and 3\% for \cref{sec:prototype_Camera}).

\begin{figure}
\centering
\begin{subfigure}[b]{0.32\linewidth}
   \includegraphics[width=\linewidth]{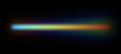}
   \caption{\footnotesize linear PSF}
   \label{fig:psf_led_linear} 
\end{subfigure}
\begin{subfigure}[b]{0.32\linewidth}
   \includegraphics[width=\linewidth]{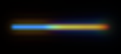}
   \caption{\footnotesize simulated learned PSF}
   \label{fig:psf_led_learned}
\end{subfigure}
\begin{subfigure}[b]{0.32\linewidth}
   \includegraphics[width=\linewidth]{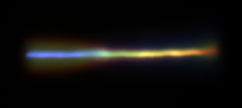}
   \caption{\footnotesize experimental PSF}
   \label{fig:psf_led_learned_real}
\end{subfigure}
\caption{\textbf{PSF coding.} The spatiotemporal PSF coding of the (a) linear focus sweep \cite{Shay2020phase}, (b) learned focus variation (simulation), and (c) same PSF in an experiment. The PSF visualizations represent the blur of a point light source moving horizontally (left to right) during the exposure time. The joint effect of the phase-mask and focus variation during exposure results in different wavelength (color) that is in-/out-of-focus when the point moves.}
\label{fig:psf}
\end{figure}

\begin{figure}
 \begin{center}
    \includegraphics[width=1\linewidth]{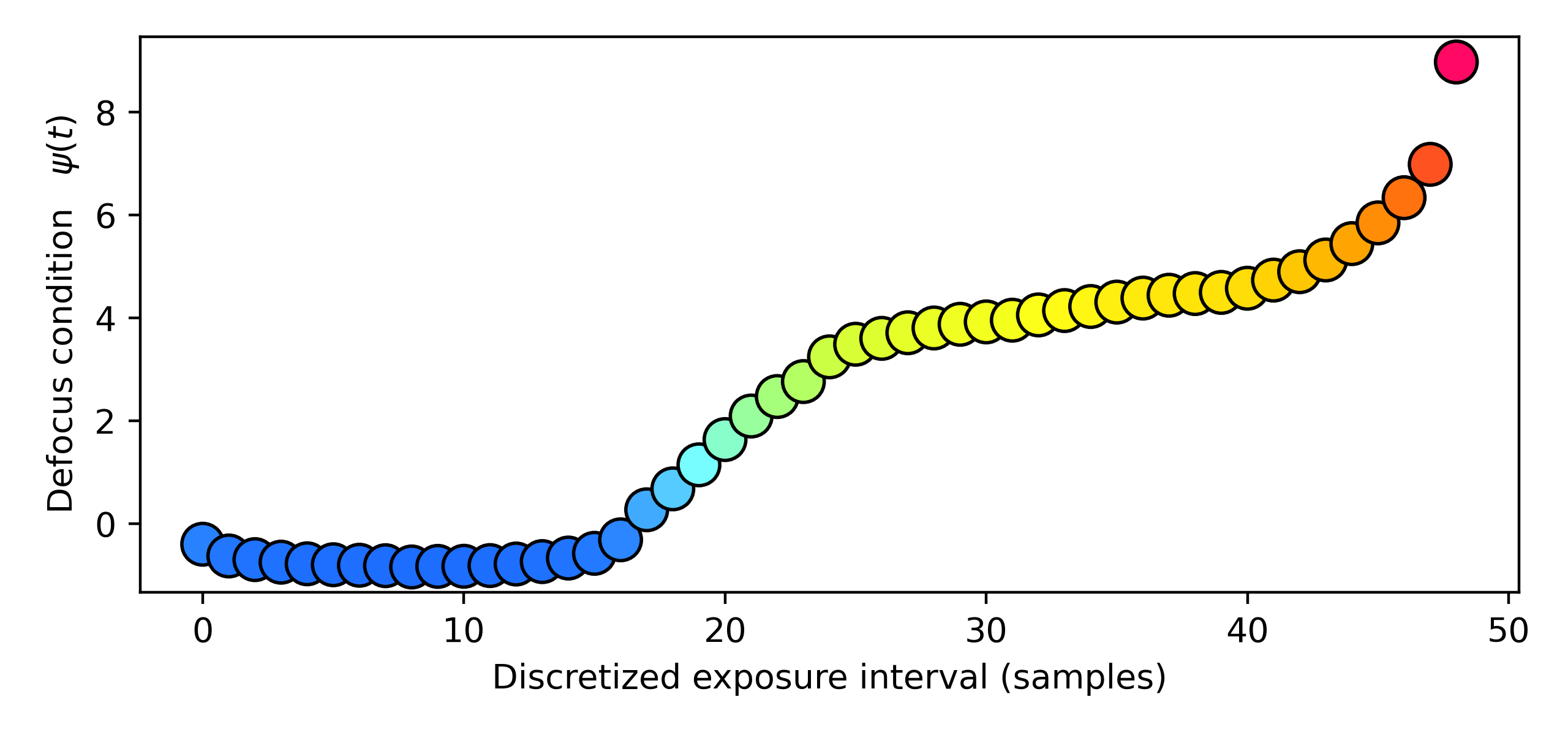}
   \caption{\op{\textbf{Learned defocus vector $\psi$.} The Learned parameters of the defocus change during the acquisition interval. The color of each sample represents the color response of the corresponding PSF kernel (at its center) as presented in \cref{fig:grid_psf_kernels}.}}
   \label{fig:psi_vector_graph}
\end{center}
\end{figure}

\op{The optimized focus sweep parameters (of the imaging simulation layer) presented in \cref{fig:psi_vector_graph}, and resulted in a coding compactly demonstrated in  \cref{fig:psf_led_learned} (the individual PSF kernels are presented in \cref{fig:grid_psf_kernels})}. In this example, a motion blur of a white dot moving right is simulated with a coding based on either linear or learned focus sweep (\cref{fig:psf_led_linear} and \ref{fig:psf_led_learned} respectively). Compared to the white trace that would have been captured in a conventional camera, the color coding of the motion profile is clearly visible. The learned pattern provides improved coding for video reconstruction, thanks to the end-to-end optimization with the image-to-video CNN. Following the different initialization methods (discussed in \cref{psf_design}) we infer that a clearly changing code with prominent characteristics is required for the reconstruction, and injective code function (such that each color appears once) helps the reconstruction and provides better results. The learned coding is also validated experimentally on a moving point source (\cref{fig:psf_led_learned_real}).

\subsection{Simulative Experiment} \label{sec:simRes}

 \begin{figure}[t] 
  \begin{center}
   \includegraphics[width=\linewidth]{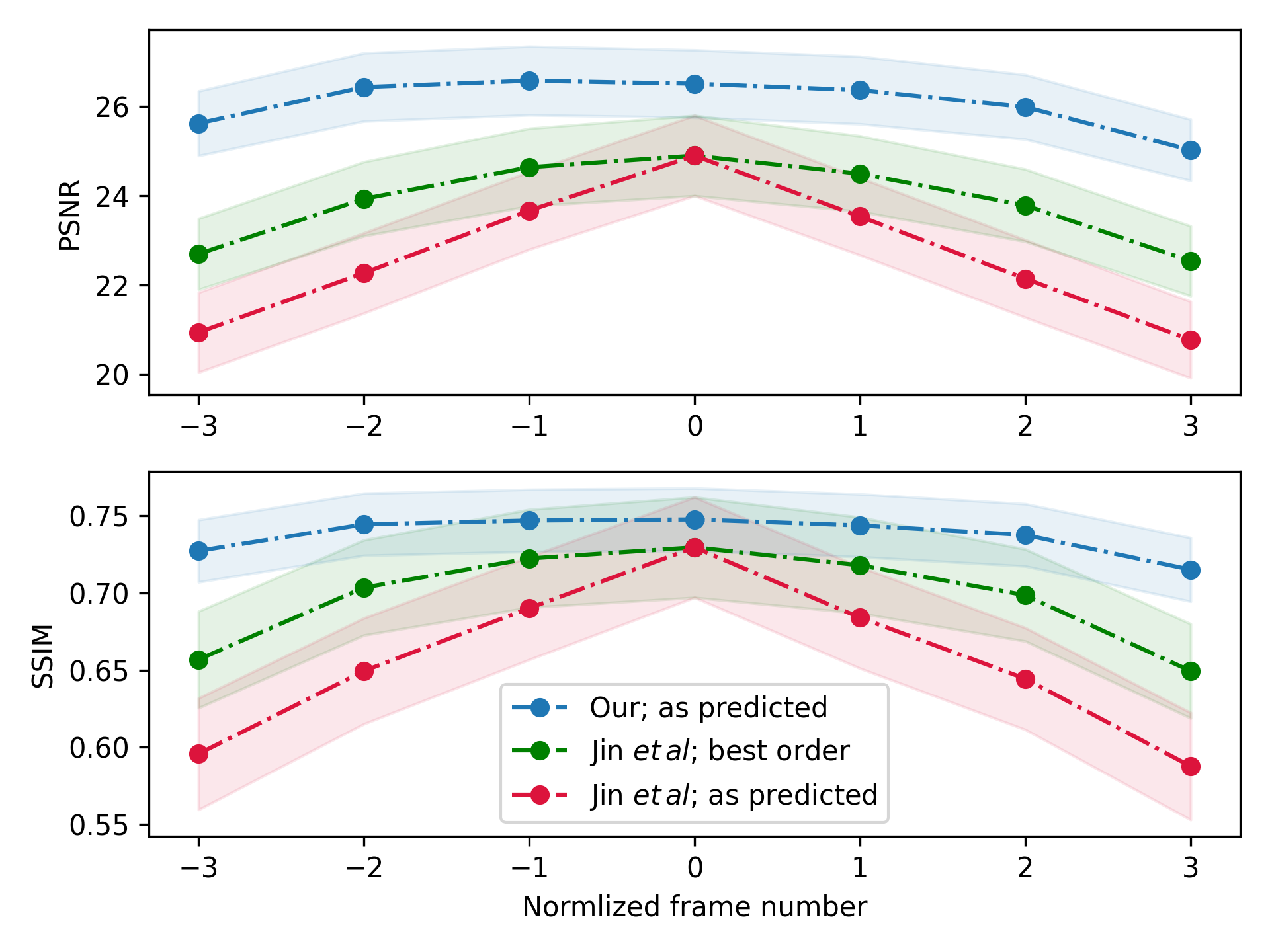}
   \caption{\textbf{Per-frame performance evaluation.} PSNR and SSIM averaged per-frame reconstruction performance for a 7-frames burst, for our method and \cite{Jin_2018_CVPR}. Since the motion blur of conventional camera is undirected, we also evaluate the reverse order of \cite{Jin_2018_CVPR} reconstructed frames (compared to the ground truth) for each input scene, and considered the higher results for the 'best order' presented evaluation.}
   \label{fig:psnr_garph}
\end{center}
\end{figure}
 \begin{figure}[t] 
 \begin{center}
   \includegraphics[width=\linewidth]{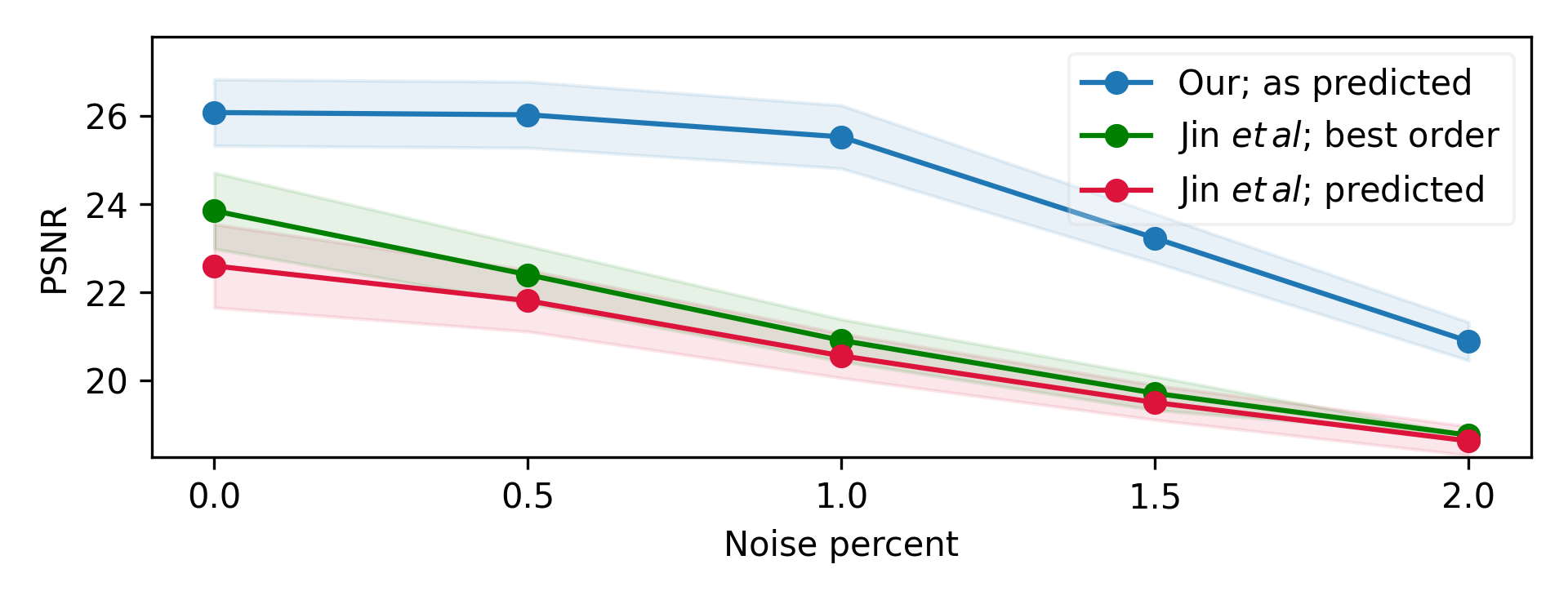}
   \caption{\textbf{Noise sensitivity analysis.} Averaged PSNR results vs. noise level (as percent of the image dynamic range) of our method and \cite{Jin_2018_CVPR} (in both predicted and best order). Our method has better noise robustness, due to the optically embedded cues.}
   \label{fig:noise_garph}
\end{center}
\end{figure}

To evaluate the reconstruction results we used a test dataset consisting of motion blurred simulated images (both conventional and coded). We compare our results to the performance of \cite{Jin_2018_CVPR} that presented a method for video reconstruction from a conventional camera (uncoded) motion blurred images.\footnote{The comparison is made only to \cite{Jin_2018_CVPR} as other related works did not publish their code for evaluation.}
We evaluate the models with respect to the GT sharp scene images using PSNR, structural similarity index measure (SSIM) \cite{SSIM} \op{and our VID metric which we designed to assess a video frame sequence reconstruction quality. 

The VID metric uses the output of the first three 3D-convolutional layers of 3D-ResNet network \cite{Tran2018ACL} (similar approach as the video loss), and is computed by taking their average in log scale (in [dB], higher is better, more details are provided in \cref{sec:Vidloss_ap}). 
Indeed, the VID metric is similar to the video loss that we use. Yet, we believe that using VID as a performance measure is far due to the following reasons: (i) We observed visually that better VID correlates with improved visual quality, which confirms the use of this loss; (ii) In the same way that it is valid to train a network using a MSE loss and report performance in terms of PSNR, it is valid to use the video loss in our training (which is not the only loss used) and report performance in terms of the VID metric.}

 \begin{figure*}[t]
 \begin{center}
    \animategraphics[loop,width=0.145\linewidth]{6}{Figs/jin_compare/legs_video/frame_}{0}{33}
    \includegraphics[width=0.85\linewidth]{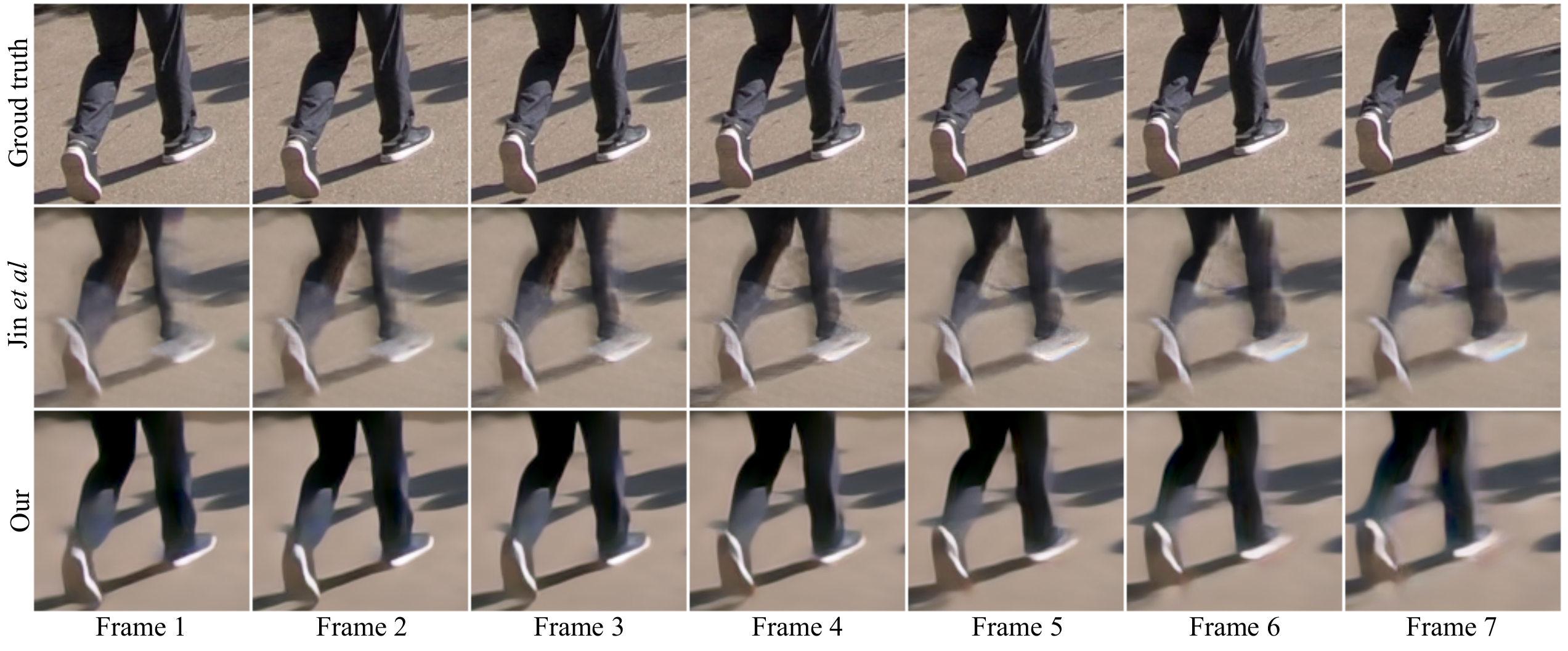}
   \caption{\textbf{Reconstruction performance (simulation).} (top row) GT image and zoom-in for a 7-frames burst, (middle row) conventional blur and \cite{Jin_2018_CVPR} results, and (bottom row) our coded input and reconstruction results. Our method achieves improved results along the entire burst and also provides a higher frame rate video. Click on the blurred input images (left) to play the result videos.}
   \label{fig:legs}
\end{center}
\end{figure*}

A visual example of our reconstruction performance is presented in \cref{fig:legs}, where improved results along the entire frame burst can be clearly seen. \Cref{fig:psnr_garph} presents the per frame performance in PSNR and SSIM (for a 7-frames burst, as \cite{Jin_2018_CVPR} is limited to such burst length only) averaged over all the test scenes. \cref{table_jin_comp} shows the overall statistics of the evaluated metrics of the reconstructions.
Since the motion direction is lost in conventional motion blur, the frames' reconstruction of \cite{Jin_2018_CVPR} may be predicted in the reversed order, i.e. in the opposite motion direction. Thus, each reconstructed scene was compared to the GT in both the predicted order and the reverse order, and the higher one (PSNR-wise) was selected to the 'best order' average. Note that in $\sim50\%$ of the cases higher performance is achieved in the reversed order, which shows that the order ambiguity is prominent.
Since the coded blur in our camera is designed to provide direction cues, our method is expected to reconstruct the frames in the correct order. Therefore, we do not need to reverse the order for it.

\begin{table}
  \centering
  \begin{tabular}{@{}lcc|cc|c@{}}
    \toprule
    & \multicolumn{2}{c|}{PSNR} & \multicolumn{2}{c|}{SSIM} & 
    \multicolumn{1}{c}{VID $\uparrow$} \\
    \toprule
    Method & mean & std & mean & std & mean\\
    \midrule
    \cite{Jin_2018_CVPR} & 22.6 &  3.78 & 0.654 & 0.143 & 10.67\\
    Best order \cite{Jin_2018_CVPR} & 23.85 &  3.45 & 0.69 & 0.128 & 10.98\\
    Ours & \textbf{26.08} &  \textbf{3.01} & \textbf{0.737} & \textbf{0.081} & 11.30\\
    \bottomrule
  \end{tabular}
  \caption{\textbf{Quantitative comparison.} \op{PSNR, SSIM and VID metrics on the entire test set. The PSNR and SSIM metrics averaged over all the reconstruction timesteps during the acquisition interval (7 timesteps), while the VID metric evaluates the whole scene sequence internally. Comparison of reconstruction quality by our method and \cite{Jin_2018_CVPR} presented both evaluation in predicted order and best order sequence (since the direction ambiguity).}}
  \label{table_jin_comp}
\end{table}

To assess the benefit in noise robustness of the encoded optical cues, a noise sensitivity analysis is carried by evaluating the reconstruction results of our method vs. the work in \cite{Jin_2018_CVPR} for different noise levels (\cref{fig:noise_garph}). Similar to the performance analysis in \cref{fig:psnr_garph}, The reconstruction performance of \cite{Jin_2018_CVPR} is evaluated both in the predicted order and best order. The prominent gap is achieved due to the optically encoded motion information, which allows reconstruction with much better noise robustness.

 \begin{figure}[t]
 \begin{center}
   \includegraphics[width=0.8\linewidth]{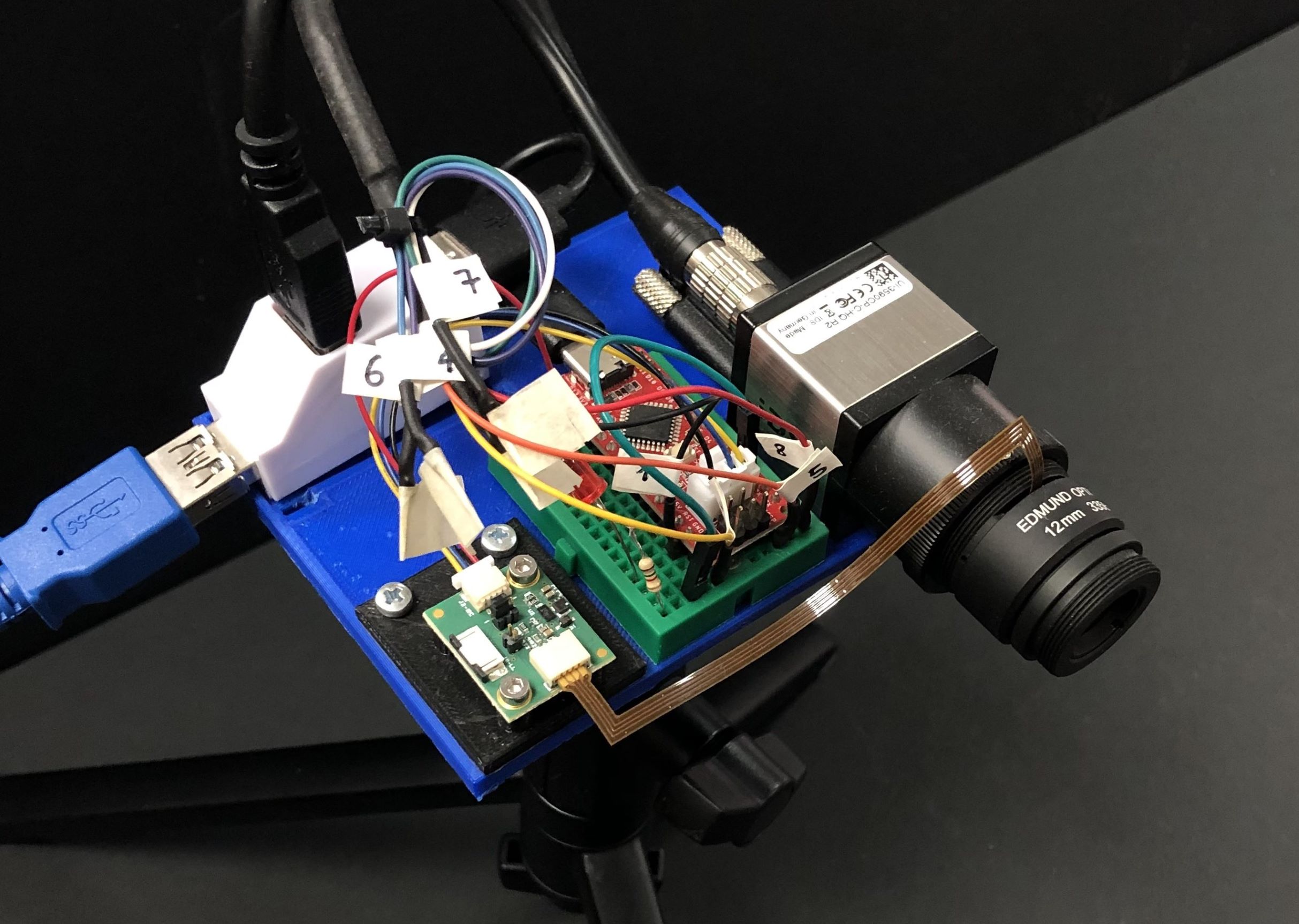}

   \caption{{\bf{Prototype Camera.}} the dynamic phase coded camera prototype is based on a commercial camera and a lens with a focusing mechanism, where our phase-mask is incorporated in the lens aperture. The camera flash signal is utilized to trigger the focus variation, controlled using the micro-controller (located near the camera).}
   \label{fig:setup}
\end{center}
\end{figure}

\subsection{Prototype Camera Results} \label{sec:prototype_Camera}
\begin{figure*}[t]

    \begin{subfigure}[b]{0.19\linewidth}
    \centering
\animategraphics[loop,width=0.95\linewidth]{6}{Figs/exp/cam_video/frame_r25_}{0}{32}
\caption{blurred input - click to play the reconstructed videos}
   \label{fig:sdfsff943} 
\end{subfigure}
    \begin{subfigure}[b]{0.8\linewidth}
    \includegraphics[width=0.98\linewidth]{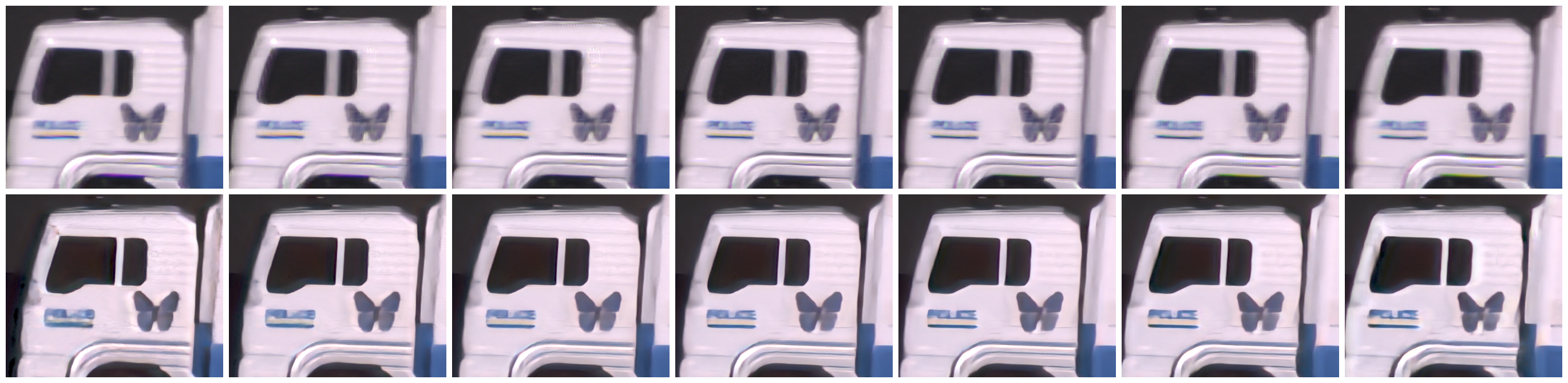}
     \includegraphics[width=0.98\linewidth]{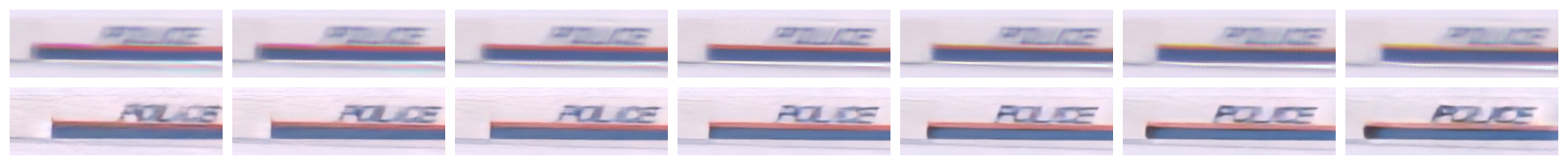}
    \centering

    \caption{Zoom-ins on 7 frames}
   \label{fig:fig1res} 
\end{subfigure}
   
   \caption{\textbf{Real-world results.} (a) blurred image from (top) conventional camera and (bottom) our coded camera; click on the blurred images to play the output videos, (b) Zoom-ins on 7 reconstructed frames of (top) \cite{Jin_2018_CVPR} and (bottom) our results. Our method achieves improved results along the entire burst, reconstructs the correct motion direction and also provides a higher frame rate video.}
   \label{fig:expRes_1}
\end{figure*}

To assess our method on real-world scenes, a prototype camera with dynamic phase coding was implemented. The color-focus phase-mask is incorporated in the lens aperture, and lens defocus setting $\psi(t)$ is set to vary during exposure following the desired learned code using a liquid-lens. The joint operation of the phase-mask and focus variation temporally manipulates the PSF $h(\psi(t))$ as presented in \cref{coded_camera_eq}. 
\op{Our prototype camera (see \cref{fig:setup,fig:blkDgrm}) is based on a standard C-mount lab-camera (IDS UI-3590CP) equipped with a 4912 x 3684 pixels ($1.25[\mu m]$ pixel pitch) color CMOS sensor \cite{UI-3590CP}. The camera is mounted with a fixed focal length $f=12[mm]$ C-mount lens with a focusing mechanism based on a liquid-lens (Edmund Cx C-mount lens \#33-632 \cite{liqlens})}
 and additional details on its design are provided in \cref{sec:Opticalsystem}. Several dynamic scenes had been captured using the prototype camera, and processed using our image-to-frames CNN for different $t$ values, thus creating short videos of the moving scenes. For comparison, we took motion-blurred images of the same scenes with a conventional camera (i.e. with constant focus and clear aperture). The results are presented in \cref{fig:teaser,fig:expRes_1}. Note how the truck moves and its back wheel rotates (front wheel is fixed) in \cref{fig:expRes_1}. Our method provides sharp results and a higher frame rate video. Note also that \cite{Jin_2018_CVPR} reconstructs the motion in the opposite direction.

\subsection{Ablation Study} \label{sec:ablation}
We apply an ablation study for the proposed method and architecture to evaluate the contribution of each component in our system. \cref{table_ablation} presents the different experiments applied and tested configurations, and \cref{miidleframe_table} presents the different coding and architectures contributions. Presented in \cref{table_ablation}, firstly we started from a UNet architecture controlled by a time parameter using AdaIN modules as described in \cref{sec:cnnArch}, where the input is the blurred image only, and the output is the sharp frame in the desired relative time in the exposure interval, and trained without the video-perceptual loss. Keeping the encoder part of the UNet uncontrolled by the time parameter (using instance normalization instead of AdaIN) enables better reconstruction results (config-a in \cref{table_ablation}) compared to the full AdaIN network, both in encoder and decoder (config-0 in \cref{table_ablation}). The following configurations include the addition of image-coordinates positional encoding features (config-b) and the time parameter concatenated to the input image (config-c). These features achieve improvement in PSNR while the similarity measure is slightly decreased, however, while testing the models on the prototype camera images we noticed better generalization to the real world images using these additions. Adding the video-frames perceptual loss (by setting $\alpha_{vid}=0.1$, see config-d) we get improvement both in PSNR and SSIM.
To comprehend the improvement of our optics and computational imaging method for the task, we train our best network (config-d) on uncoded images (i.e. temporal averaging only), and evaluated the results (config-e). Without the phase coding we observe a significant performance degradation, which validates the optical coding benefit to the reconstruction ability. 
Using the learned temporal coding we gain improvement in both reconstruction metrics (config-f), and we consider it as our proposed model. \op{From the VID metric evaluation, which represents the video reconstruction perceptuality, we observe significant improvement using our learned coding compare to the linear (config-f and config-d respectively)} 

\begin{figure}[t]
 \begin{center}
   \includegraphics[width=\linewidth]{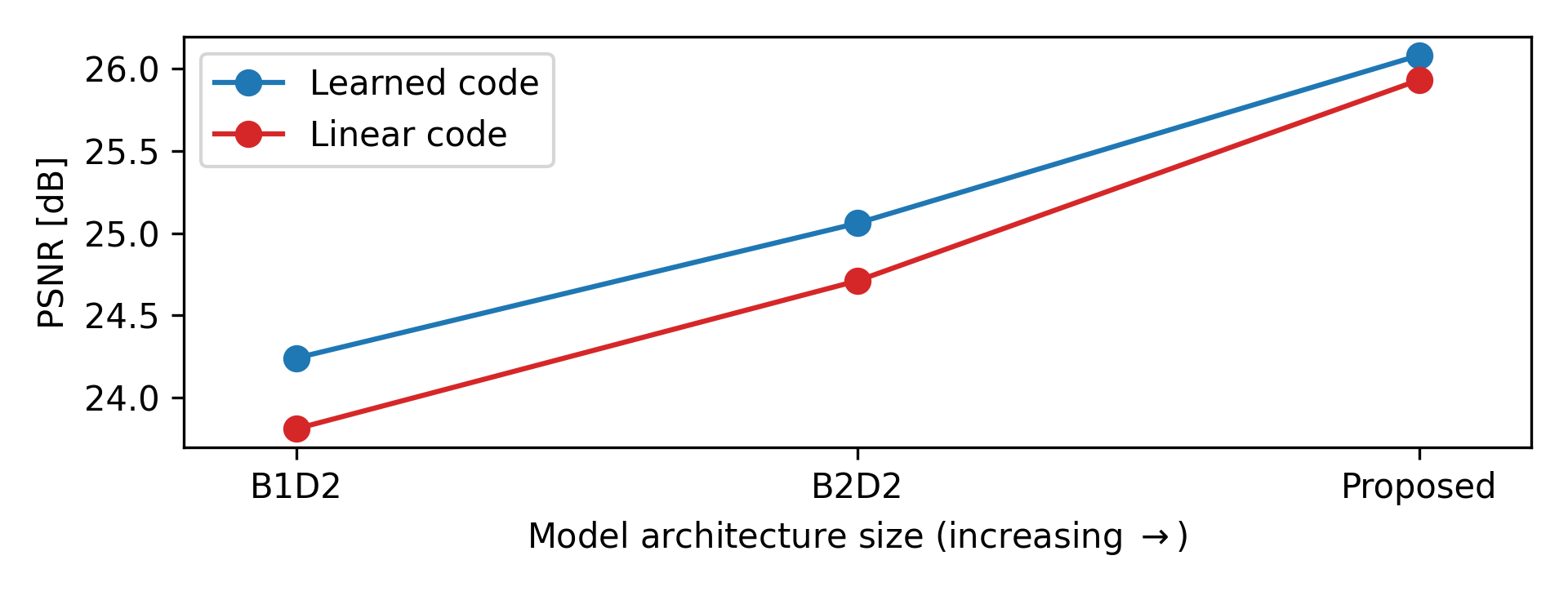}

   \caption{\op{\bf{Reconstruction PSNR vs. model size.} PSNR reconstruction results for three different size models: Our proposed model and two lighter Unet models with encoding depth of two and double/single convolutions block (B2D2 and B1D2 respectively, details in \cref{sec:additional_models}). Our learned code improvement is more significant as the model is more degraded and less powerful.}}
   \label{fig:model_size_graph}
\end{center}
\end{figure}

\op{To comprehend better the improvement achieved by our learned code (compare to the linear coding) we evaluate different size models as presented in \cref{fig:model_size_graph}. Using smaller models (e.g. in limited resources conditions) the improvement of the learned code becomes more significant and it contributes to better reconstruction results of the degraded models. We conclude that while a large network is capable to solve the harder task (linear code), for a smaller (and weaker) network the learned code is more meaningful to achieve better results. For the smaller networks, we used U-net with encoding depth of two levels and Mobile-Net blocks \cite{mobilenet} (additional details about the models' architecture described in \cref{sec:additional_models}).}

In \cref{miidleframe_table} we evaluate the coding methods contribution by the central frame performance (namely the deblurring performance). The uncoded exposure using \cite{Jin_2018_CVPR} reconstruction method achieves inferior results. The performance of the naive UNet architecture with linear coding is presented as well. The suggested model is also presented in the table. It is noticeable that the improved UNet improves the results while the learned PSF achieves additional improvement. \op{We also trained our Unet model only on the central frame (as a deblurring task) to infer about the flexibility-quality tradeoff. we used the linear code (equivalent to \cite{Shay2020phase}) and our learned code. Even though our proposed method suffers from a small performance drop on the middle frame, it allows the flexibility of generating a video sequence of the scene with any desired number of frames. For these models we replaced AdaIN with group normalization (GN) for better stability (additional details in \cref{sec:additional_models}). We also present another variant of our model using AdaGN instead of AdaIN (config-g in \cref{table_ablation}), such a model achieves better results for PSNR and SSIM, but similar results in VID metric (architecture details in \cref{sec:additional_models}. Even though, we consider config-f in \cref{table_ablation} as our suggested model for the comparisons and the obtained visual results.}

\begin{table}
  \centering
  \begin{tabular}{@{}llccc@{}}
    \toprule
    
    &Method & PSNR $\uparrow$ & SSIM $\uparrow$ & VID $\uparrow$\\
    \midrule
        (0)&time dependent encoder & 21.5 & 0.69 & 10.59 \\
    (a)&UNet & 25.06 & 0.73 & 11.07\\
    (b)& + positional encoding & 25.16 & 0.728 & 11.08 \\
    (c)& + time concatenation & 25.70 & 0.705 & 11.15\\
    (d)& + video perceptual loss & 25.93 &  0.735 & 11.12 \\
    (e)& (d) w/o phase coding & 22.96 & 0.645 & 10.57\\
    (f)& (d) with learned PSF & 26.08 & 0.737 & \textbf{11.30}\\
    (g)& +AdaGN instead of AdaIN & \textbf{26.38} & \textbf{0.75} & \textbf{11.30}\\

    \bottomrule
  \end{tabular}
  \caption{\textbf{Ablation study.} To assess the contribution of each feature of our method, we performed a gradual performance evaluation. \op{The PSNR and SSIM metrics were averaged over all the reconstruction timesteps during the acquisition interval, while the VID metric evaluates the whole scene sequence internally. All the metrics were averaged over the test set scenes.}}
  \label{table_ablation}
\end{table}

\begin{table}
  \centering
  \begin{tabular}{@{}lccc@{}}
    \toprule
    Method & PSNR $\uparrow$ & SSIM $\uparrow$\\
    \midrule
    Uncoded \cite{Jin_2018_CVPR} & 24.9 & 0.729 \\
    UNet + linear code (a) & 25.48 & 0.74 \\
    Suggested UNet + linear code (d) & 26.3 & 0.744 \\
    Suggested UNet + learned code (f) & 26.51 & 0.747 \\
    \op{Suggested UNet + AdaGN + learned code (g)} & \textbf{26.75} & \textbf{0.759} \\
    \midrule
    \multicolumn{3}{@{}l}{\op{Deblurring models (trained only on the central frame)}} \\ 
    \hspace{3mm} Linear coed - central frame with GN & 26.73 & 0.745 \\
    \hspace{3mm} Learned coed - central frame with GN & \textbf{27.05} & 0.757 \\
    \bottomrule
  \end{tabular}
  \caption{\textbf{ Central Frame Performance.} Averaged PSNR/SSIM metrics on the
central frame (on the test dataset) for different coding methods: uncoded, linear and learned. (the letter in the parenthesis indicates the entry in \cref{table_ablation}). \op{We also present models trained only for the deblurring task (central frame reconstruction) for flexibility-quality tradeoff assessment.}}
  \label{miidleframe_table}
\end{table}

\begin{figure}[t]
 \begin{center}
    \includegraphics[width=1\linewidth]{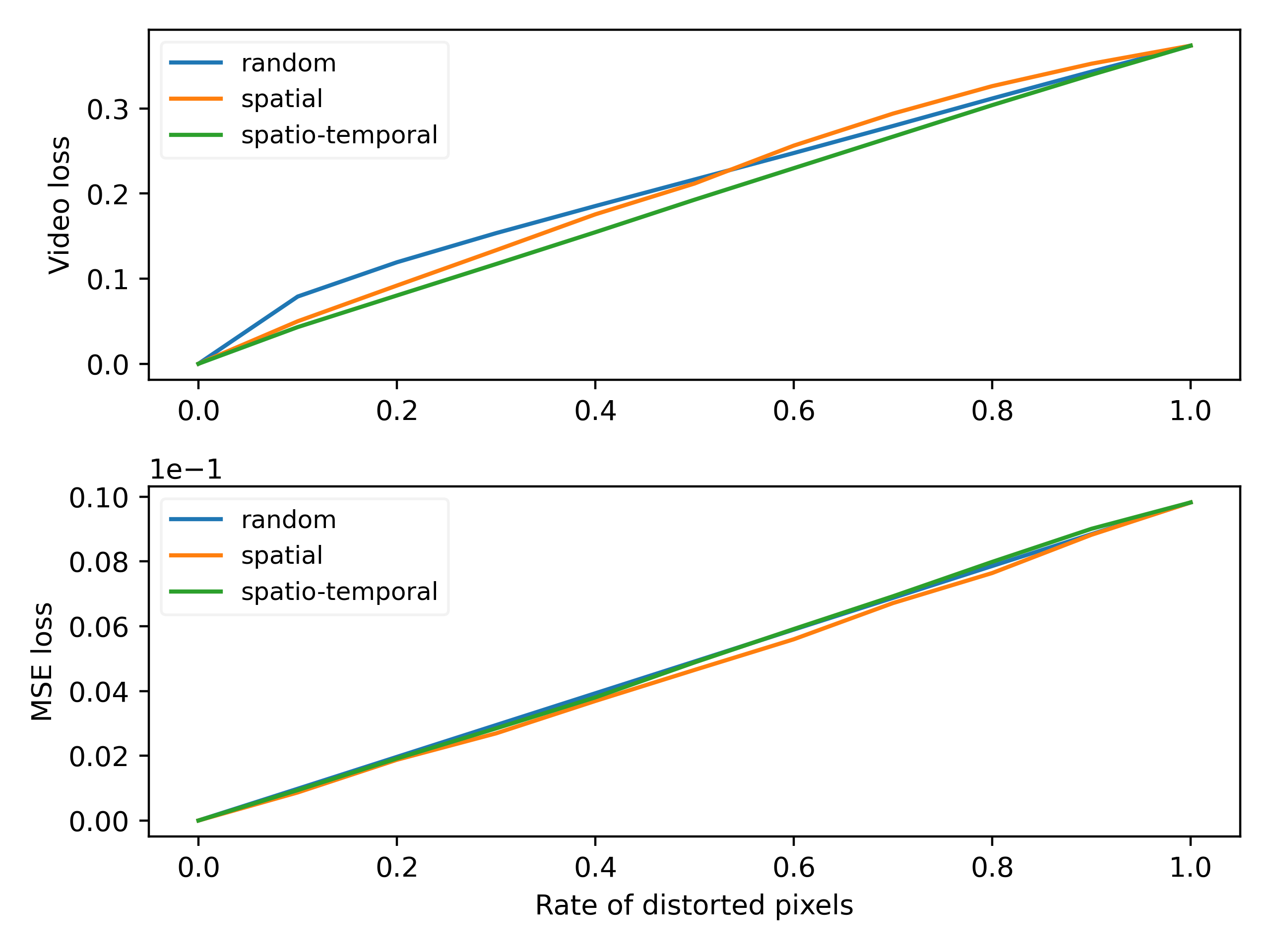}
   \caption{
   \opb{
   \textbf{Video loss validation by consistent distortions.} For the VID loss it is noticeable that the spatio-temporal consistency distortions achieve the lower loss for all distortions rates, and the loss is affected by video consistency. On the other hand, MSE loss has no spatial correlation and thus is not affected by the spatio-temporal consistency.
   }
   }
   \label{fig:vidloss_exp1}
\end{center}
\end{figure}

\opb{
\subsection{VID Loss/Metric Validation} \label{sec:videoloss_details}
In this section we validate the proposed video loss and present the spatial and temporal consistency preserving behavior of the metric. Since the loss is based on a pretrained 3D-ResNet architecture, the explainability of the model is a challenging task. The video loss is reference-based, namely, it requires the ground-truth video along with the distorted (reconstructed) video to compare the deep features difference. The well known MSE loss is also reference-based in this sense, but it compares the pixel values pixel-wise. The MSE loss has neither spatial nor temporal dependencies between pixels, unlike our video loss. 

The first observation is that both losses typically decrease or increase together depending on the similarity between the distorted and ground-truth videos. To present the spatio-temporal consistency of the video loss we use the following test scheme: we choose a fraction $p$ of pixels of the video to be distorted. Note that for $p=1$ we get the worst loss value since all the video pixels are distorted, and for $p=0$ we get zero loss (identical to GT). We choose the pixels to distort in three methods:  (i)random over space and time; (ii) not altering some spatial blocks that preserve spatial consistency (a rectangle in each frame randomly located over the time axis); and (iii) not altering some spatial-temporal blocks that preserves both a consistency in time and space (rectangle in each frame in the same location over time axis). The results are presented in \cref{fig:vidloss_exp1}. It is noticeable that for the spatio-temporal consistent distorted video the video loss gives the lowest results (for each $p$ value). Due to MSE loss's lack of spatial and temporal dependence, the metric behaves the same regardless of the pixel sampling (and distortion) method.

For an additional observation, we fixed $p=0.5$ and tested the temporal and spatial consistency for two video distortion types: spatial Gaussian blur ($\sigma=1$) and pixels shift (3 pixels in both axes). For the inconsistent temporal/spatial case, we set every other frame/row to be distorted, while for the consistent case we set the first half of the frames/rows to be distorted. As presented in \cref{vidloss_consist_table}, the VID loss is much better in the consistent case (for both spatial and temporal), while the MSE loss is not affected much due to a lack of axial correlation. Hence, the VID loss encourages spatial and temporal consistency as a training loss. 
}

\begin{table}
  \centering
  \begin{tabular}{@{}lcc|cc|cc@{}}
    \toprule
    Metric & \multicolumn{2}{|c}{Inconsistent} & \multicolumn{2}{|c}{Consistent} & \multicolumn{2}{|c}{Loss change}\\
    \midrule
    Loss & spat & temp & spat & temp & spat & temp\\
    \midrule
    \multicolumn{3}{@{}l}{Frame blur:} & \multicolumn{2}{|c}{} & \multicolumn{2}{|c}{} \\ 
    \hspace{2mm} MSE ($^{-4}$) & 6.65&6.68 & 6.9&6.68 & +3.53\% &$\sim$0\% \\
    \hspace{2mm} VID ($^{-2}$) & 3.92 & 4.49 & 3.38 & 4.1 & \textbf{-16.1\%}&\textbf{-10.2\%} \\
    \multicolumn{3}{@{}l}{Frame shift:} & \multicolumn{2}{|c}{} & \multicolumn{2}{|c}{}\\ 
    \hspace{2mm} MSE ($^{-3}$) &8.22 &8.21 & 8.54 &8.19 & +3.83\%&$\sim$0\% \\
    \hspace{2mm} VID ($^{-2}$) &10.9 &10 & 7.36 &8.4 & \textbf{-47.7\%} & \textbf{-18.4\%} \\
    \bottomrule
  \end{tabular}
  \caption{\opb{\textbf{Consistency significance for VID loss} we tested the spatial/temporal consistency (denoted as spat./temp. respectively) effect on the VID and MSE losses (for $p=0.5$). The VID loss is very affected by the consistency of the distorted data, and thus encourages the models for consistent predictions. (the loss values are in $10^x$ scale according to the value in the brackets of each line)}
  }
  \label{vidloss_consist_table}
\end{table}

\begin{figure}
\centering
    \includegraphics[width=1\linewidth]{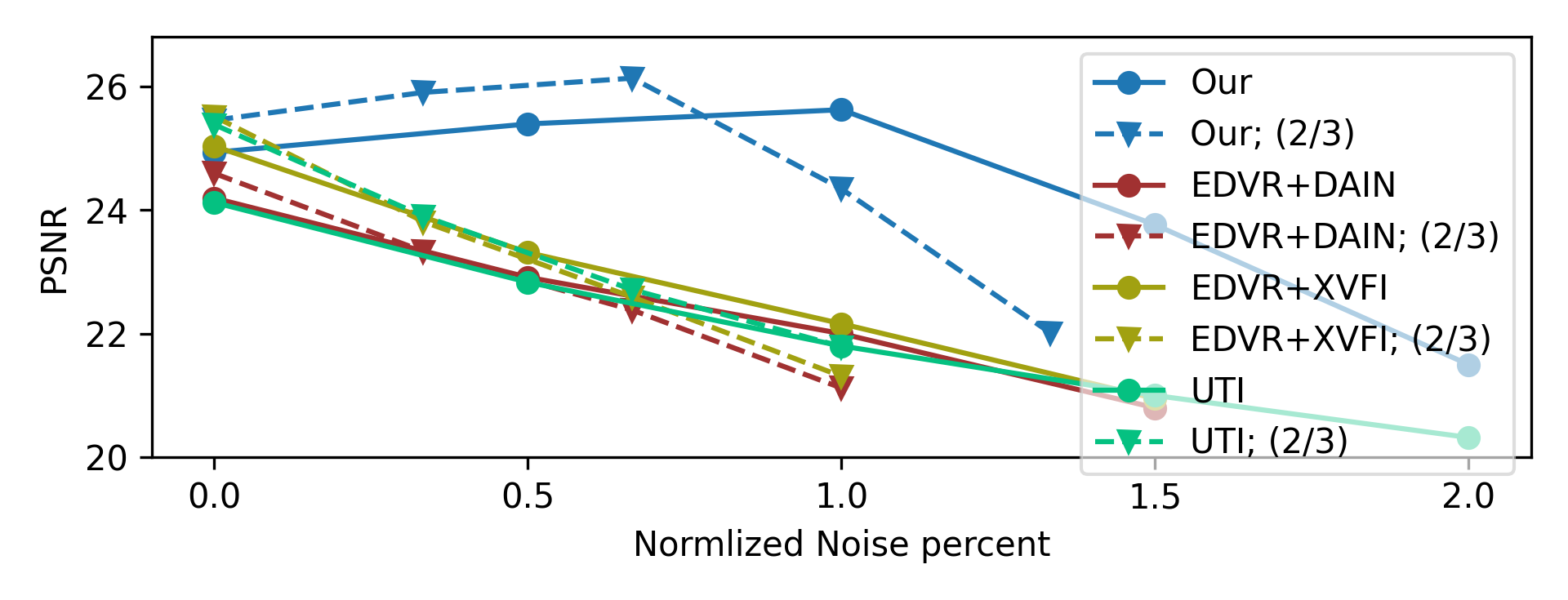}
\caption{\opb{\textbf{Video frame interpolation - PSNR performance for different noise levels on REDS dataset \cite{Nah_2019_CVPR_Workshops}} Comparison of PSNR for two exposure intervals: baseline and two-thirds of the baseline. The noise axis was normalized with respect to the exposure interval based on the SNR behavior (more information in \cref{SNRtradeoff}).
}}
\label{fig:vfi_psnr_reds}
\end{figure}

\begin{figure}[t]
 \begin{center}
    \includegraphics[width=1\linewidth]{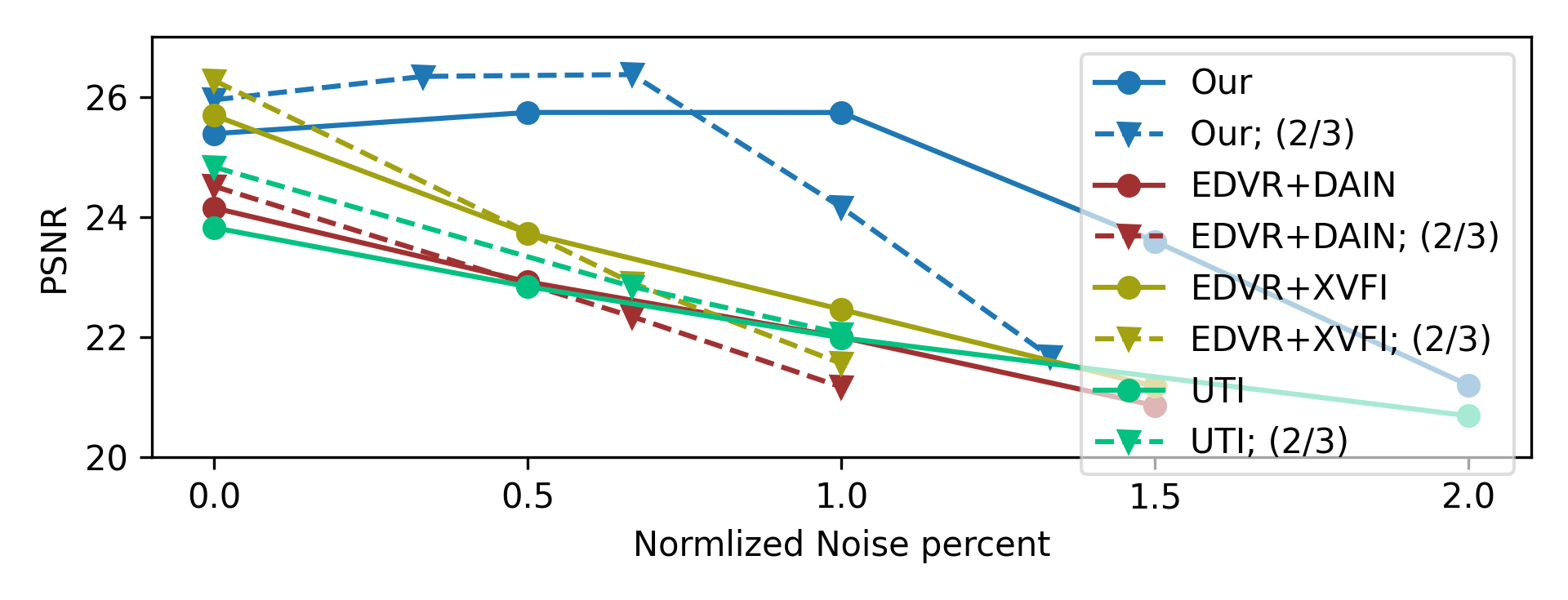}
   \caption{
   \opb{
   \textbf{Video frame interpolation - PSNR performance for different noise levels on Adobe240 dataset \cite{su2017deep_adobe240}.} Comparison of PSNR for two exposure intervals: baseline and two-thirds of the baseline. The noise axis was normalized with respect to the exposure interval based on the SNR behavior (more information in \cref{SNRtradeoff}).
   }
   }
   \label{fig:vfi_psnr_adobe}
\end{center}
\end{figure}

\begin{figure}
\centering
    \includegraphics[width=1\linewidth]{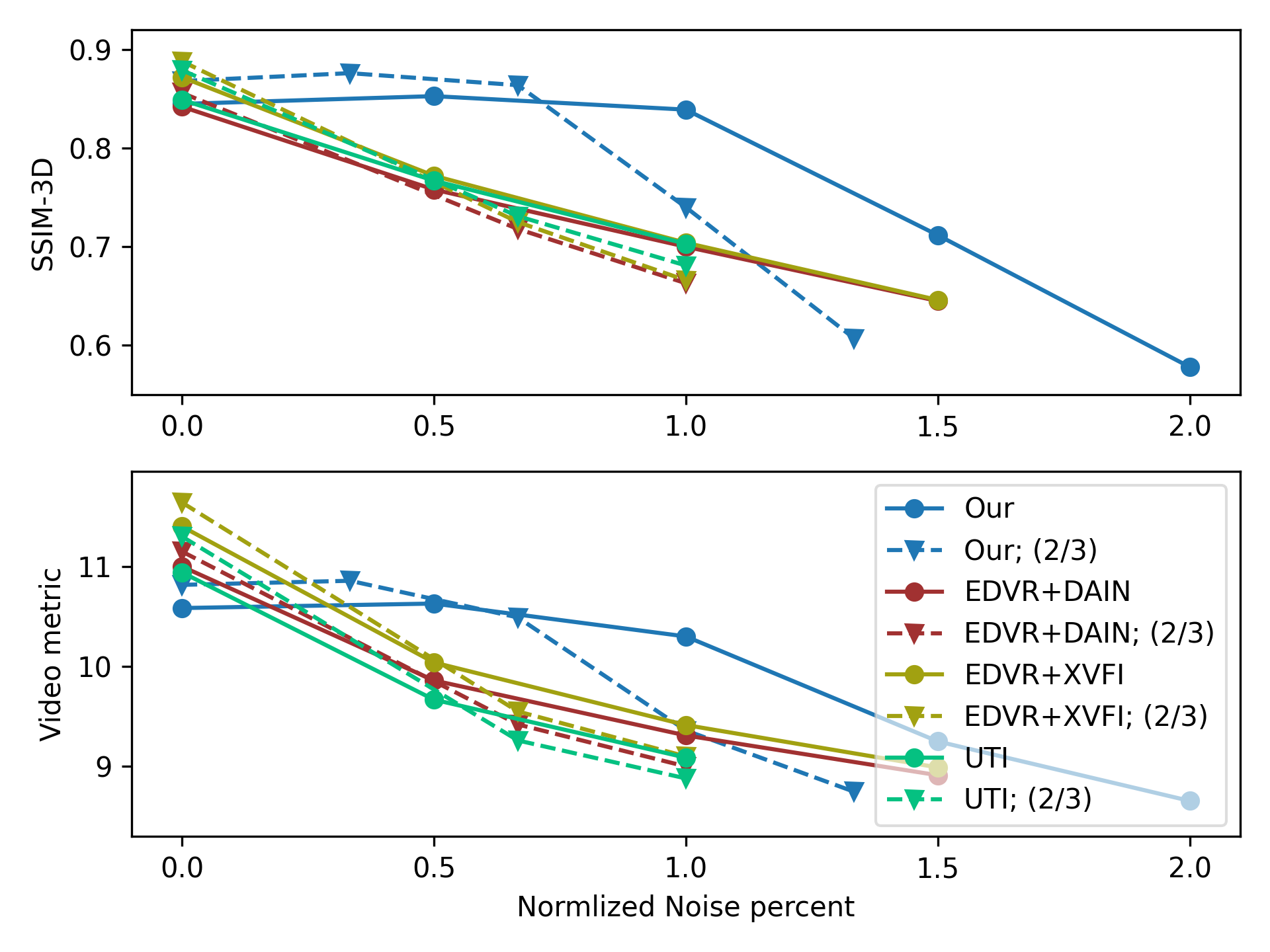}
\caption{\opb{\textbf{Video frame interpolation - SSIM3D and VID performance for different noise levels on Adobe dataset \cite{su2017deep_adobe240}.} SSIM-3D and VID performance for two exposure intervals: baseline and two-thirds of the baseline. The noise axis was normalized with respect to the exposure interval based on the SNR behavior (more information in \cref{SNRtradeoff}).
}}
\label{fig:vfi_ssim_adobe}
\end{figure}

\opb{\subsection{Video Frame Interpolation Extension} \label{sec:vfi} 
We extend our encoding method for video frame interpolation. Different from the single image task, in this case we use consecutive blurred video frames as input and perform both deblurring and video frame interpolation. Previous works suggested methods for this task, such as  \cite{Shen_2020BIN_CVPR, NEURIPS2020_UTI}, while other works designed solutions for interpolation of sharp input video frames, as \cite{DAIN, Sim_2021XVFI_ICCV}.
In our method, each frame in the input video is encoded using our learned spatiotemporal PSF. For the sharp frame interpolation methods \cite{DAIN, Sim_2021XVFI_ICCV}, we first applied a video deblurring algorithm using \cite{wang2019edvr}. We compare the results of the mentioned methods on REDS dataset and Adobe240 dataset \cite{su2017deep_adobe240}, which is a different domain from the data we used for training and validation. We synthesized blurred frames in two timing setups, a baseline exposure and two-thirds of the baseline (see \cref{sec:vfi_details}). The performances of the methods were evaluated as a function of noise levels by PSNR (\cref{fig:vfi_psnr_reds} and \ref{fig:vfi_psnr_adobe}) and SSIM-3D and VID (\cref{fig:vfi_ssim_adobe} and \ref{fig:vfi_ssim_reds}). It is noticeable that our method is more robust and performs better for noisy images, while some of the other methods perform better for clean images which are impractical for real camera images.  
We also test \cite{Shen_2020BIN_CVPR} method but omit it from the graphs due to low performance (4dB PSNR lower than other methods). Note that since we optimized our model for noise with $\sigma=1$ there is a slight drop in performance for lower noise values. In \cref{sec:vfi_details} we elaborate on training details and present additional results for REDS dataset. Visual results of the reconstructed videos are presented in \cref{fig:vfi_results_n05} and in the supplementary material.
}

\opb{\subsection{SNR Tradeoff In Dynamic Scenes}
\label{SNRtradeoff}
While capturing dynamic scenes there is a tradeoff between the signal intensity and motion blur. For lower noise, a long exposure is required and thus a long motion trajectory is obtained in the captured image. On the other hand, a sharp image is obtained by short exposure but the signal level is low, and we get low SNR. Hence, there is a tradeoff between the two, and capturing images/frames without noise is impractical. Thus it is important to have noise robust method for the frame interpolation task. In the noise evaluation figures for the frame interpolation task (\ref{fig:vfi_psnr_reds}, \ref{fig:vfi_psnr_adobe}, \ref{fig:vfi_ssim_adobe} and \ref{fig:vfi_ssim_reds}), we scaled the noise levels axis of the "two-thirds" timing setup data compared to the baseline noise levels for the SNR tradeoff evaluation. Namely, since the exposure time in this timing setup is two-thirds of the baseline setup the motion blur is lower and the noise should be higher according to the tradeoff. Such that adding $1.5\%$ noise to short (two-thirds) exposure frames should be compared to $1\%$ noise added to the baseline exposure frames. From the results presented in the mentioned figures we can conclude that for low noise levels (brighter scenes) it will be preferred to use shorter exposure (and get sharper images) while for high noise levels it will be preferred to use longer exposures.}

\noindent {\bf Limitations.} Despite the improved performance achieved, our method still suffers from several limitations. The most prominent are scenes with high-speed or accelerating objects; as our coding method is a composition of the dynamic phase coding and object movement, there is some hidden assumption that this movement (and specifically its acceleration) is not too acute. In such cases, the resulting coded information will be too obscure, with a limited benefit. Another limitation relates to the imaging scenario; the temporal part of the coding is focus variation. Therefore, the underlying assumption in such a design is that the entire scene is in the same focus condition (either in- or out-of-focus). Such a design limits our solution to infinite-conjugate lenses (e.g. GoPro cameras). \op{This limitation is more prominent in outdoor scenes with depth where the image is not in the same focus condition.} In addition, since our coding is color-based, we assume that objects are not monochromatic, since in such a case the coding ability is degraded. This assumption is acceptable since almost all natural materials are not monochromatic, and practically even some wavelength bandwidth can suffice. Textures are important to indicate a motion in general, and the textures are required to achieve the coded blur in the image for the reconstruction. Even though, if there are no textures the reconstruction becomes a quite trivial task (due to a single color object, where motion and blur are less apparent). \op{Minor artifacts of the model might be observed by a careful analysis of the result videos.} \opb{The reconstructed result might get smooth since the blurring process may deteriorate the small details in the image, which the model may struggle to recover. Worth noting that since our training dataset is generated using hand-held camera video, there is a camera movement in the synthesized motion blur, and for such small movements, the model learns to reconstruct the motion correctly.}
\section{Conclusion}
\label{sec:summ}
\cite{Raskar2006CodedEP}

A spatiotemporally coded camera for video reconstruction from motion blur is proposed and analyzed. Motivated by the ongoing requirement to improve the imaging capabilities of cameras, the motion blur limitation is utilized as an advantage, to encode motion cues allowing reconstruction of a frame burst from a single coded image. The coding process is performed using a phase-mask and a learnable focus variation, resulting in color-motion cues encoding in the acquired image. This image, along with a relative time parameter $t$, are fed to a CNN trained to reconstruct a sharp frame at time $t$ in the exposure time. By choosing a sequence of $t$ values, a frame burst of the scene is reconstructed. Simulation and real-world results are presented, with improved performance compared to existing methods based on conventional imaging, both in reconstruction performance and handling the inherent direction ambiguity. 

\op{Moreover, we present a vast ablation study, noise robustness analysis, learned code contribution (including model size dependency), and central frame performance with a flexibility-quality tradeoff assessment.}

Our method can assist balancing the various trade-offs that a camera designer has to handle. For example, the promising results achieved hold the potential to extend the method to perform a low-blurred to high-sharp frame rate conversion, achieved with a lower sampling rate and improved light efficiency. This may extend existing photography capabilities with simple and minor hardware changes.

{\small
\bibliographystyle{ieee_fullname}
\bibliography{egbib}

}
\appendix

\op{\section{PSF computation}
\label{sec:PSFcomputation}
The PSF of the camera which is conditioned on the time varying defocus parameter (denoted as $h(\psi(t))$) is computed numerically to simulate the camera acquisition process (\cref{method_camera_dpc}), and for the end-to-end training of the defocus condition parameters vector $\overline{\psi}\in\mathbb{R}^{49}$ (\cref{sec:exp}). The computation is performed according to Fourier optics, as used also in \cite{Depth_2018,EDOF_DL}.
In out-of-focus imaging the $\psi$ defocus measure is defined as:

\begin{equation}
\label{app_psf_comp1}
\psi = \frac{\pi R^2}{\lambda} \left(\frac{1}{z_o}+\frac{1}{z_{img}}-\frac{1}{f}\right) = \frac{\pi R^2}{\lambda} \left(\frac{1}{z_{img}}-\frac{1}{z_i}\right),
\end{equation}
where $f$ is the focal length, R is the exit
pupil radius, $\lambda$ is the illumination wavelength, $z_{img}$ is the sensor plane, and $z_i$ is the ideal image plane for an object located at $z_o$. The in-focus circular pupil function is denoated as $P(\rho, \theta)$. By adding a coded pattern (amplitude,
phase or both) at the exit pupil, the PSF of the system can be manipulated by a pre-designed pattern. The coding phase mask located at the aperture is denoted as $C(\rho, \theta)$, which is a circularly symmetric piece-wise constant function representing the mask phase shift rings. Such that, for each ring $k$ between $r_{k1}<\rho< r_{k2}$ it holds that $C(\rho, \theta) = exp\{j\phi_k\}$ where $\phi_k$ is the phase shift of the ring. The specific parameters presented in \cref{sec:Opticalsystem}
The defocus parameter $\psi$ measures the maximum quadratic phase error at the aperture edge, such that we get:
\begin{equation}
\label{app_psf_comp2}
P_C(\psi) = P(\rho, \theta) \cdot C(\rho, \theta) \cdot exp(j\psi \rho^2)
\end{equation}

Following \cite{GoodmanFO}, the PSF of an incoherent imaging system is defined as:
\begin{equation}
\label{app_psf_comp3}
h(\psi) = \left|\mathcal{F}\{P_C(\rho, \theta \,;\; \psi)\}\right|^2,
\end{equation}
where $\mathcal{F}$ denote Fourier transform. 

\opb{We compute the PSF for the RGB colors ($\lambda\in \{610, 535, 455\}nm$) to simulate the camera acquisition using RGB images by Eq.~\eqref{coded_camera_eq}. It is an approximation of the real imaging system which applies a specific PSF for each wavelength in the full spectrum of light. Under the assumption that the PSF changes slowly in $\lambda$ compared to the bandwidth of the color filter array of the camera (Bayer filter, for each color of RGB) the approximation holds. 

The PSF is a two-dimensional continuous function in the spatial coordinates of the image. Due to the focus change during the exposure $\psi$ changes and thus the PSF changes continuously in time. for the acquisition simulation by Eq.~\eqref{coded_camera_eq} the spatio-temporal PSF was discretized in time and space as presented in \cref{fig:grid_psf_kernels}}.

\section{Video loss and video metric}
\label{sec:Vidloss_ap}
For a video loss and video metric (for train and evaluation respectively) we used 18-layers ResNet3D model \cite{Tran2018ACL} that performs 3D spatiotemporal convolutions on video time-space volume. We use the outputs of the three first convolution layers (namely \texttt{conv1}, \texttt{conv2} and \texttt{conv3}) for the reconstructed frame sequence and the ground truth frames (consist of 7 frames each as described in \cref{sec:cnnArch}). We compute Smooth L1 loss between the two features for each layer output, such that we get three scalar values. These values (denoted as $l_k$) represent the similarity both in spatial and temporal dimensions. For training, we average the three components to a single loss value (denoted as $l_{vid}$ in \cref{sec:cnnArch}). For the video sequence evaluation using the video metric, we compute each of the three components in log scale (same as PSNR) and averaged them to a single value (denoted as \textbf{VID} \cref{sec:cnnArch})
\begin{equation}
\label{vidmet_log}
VID_k = -10 \cdot \log_{10}l_k
\end{equation}
\begin{equation}
\label{eq:vid_loss_avg}
VID = mean(VID_1, VID_2, VID_3)
\end{equation}}

\section{Dynamic phase-coded camera prototype}
\label{sec:Opticalsystem}

Our method is based on a dynamic phase coded camera, designed to embed color-motion cues in the intermediate image, and a corresponding CNN trained to decode these cues and reconstruct a sharp frame burst. After achieving satisfying simulation results (i.e. with simulated coded images), we assembled a prototype camera implementing our proposed dynamic phase-coding. As mentioned in the paper, our coding method is relatively simple and based mostly on conventional commercial parts. As such, it can be easily integrated into any camera equipped with a focusing mechanism.

The coding is achieved jointly using a phase-mask in the lens aperture and by performing a focus sweep during the exposure time. Following the work in \cite{Shay2020phase}, we use a similar phase-mask, comprised of two phase rings. The phase-mask aperture diameter is $D=2.3[mm]$; the first ring (inner-to-outer) radii are $r=[0.633  ~ 0.92]mm$ and its phase shift is $\phi=6.5[rad]$; the second ring radii are $r=[0.92 ~ 1.15]mm$ and its phase shift is $\phi=13.2[rad]$ (the phase-shifts are measured with respect to $\lambda=455 [nm]$, which is the peak wavelength of the camera's blue channel). The phase-mask is fabricated using conventional photo-lithography and wet etching process. 

The dynamic PSF encoding is achieved by applying the learned focus variation during the exposure time. The focus change is performed electronically, using the camera focusing mechanism, controlled by a dedicated micro-controller \cite{ArduinoNano}. The micro-controller contains the learned focus sweep parameters, and triggers the required coding in synchronization with the exposure (utilizing the camera flash-signal, designed to indicate the start of the exposure). Note that although various components had been used in our implementation, the coding can also be implemented easily on existing cameras (assuming the availability of API to the focusing and exposure mechanisms).

 \begin{figure}[t]
 \begin{center}
   \includegraphics[width=0.9\linewidth]{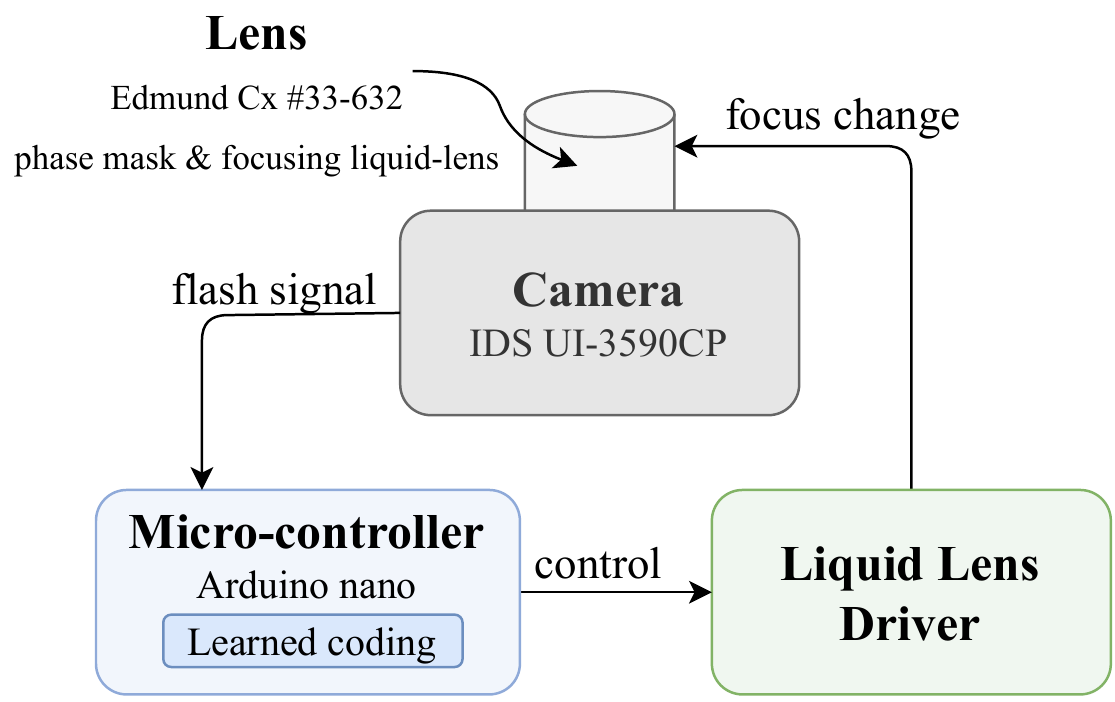}

   \caption{{\bf{Prototype Camera Diagram.}} The flash signal from the camera initiates the learned focus variation during the exposure using a micro-controller, such that the designed dynamic phase coding is performed and a motion-coded image is acquired.}
   \label{fig:blkDgrm}
\end{center}
\end{figure}


\op{\section{Learned code and PSF kernels visualization} \label{sec:codevisual}
A visualization of a single horizontally moving point light source is presented in \cref{fig:psf_led_learned}. It is achieved from \cref{coded_camera_eq} with the time changing camera PSF, denoted as $h(\psi(t))$. The vector $\overline{\psi}\in\mathbb{R}^{49}$ values (exposure time discretized for 49 values) presented in \cref{fig:psi_vector_graph}, and the resulting 49 spatial color kernels presented in \cref{fig:grid_psf_kernels} computed as presented in \cref{sec:PSFcomputation}. Note that the color at the center of each kernel was used for the $h(\psi(t))$ points in \cref{fig:psi_vector_graph}. The numerical values of the learned $\psi$ vector are provided in the additional materials with the submitted code.

 \begin{figure}[t]
 \begin{center}
    \includegraphics[width=1\linewidth]{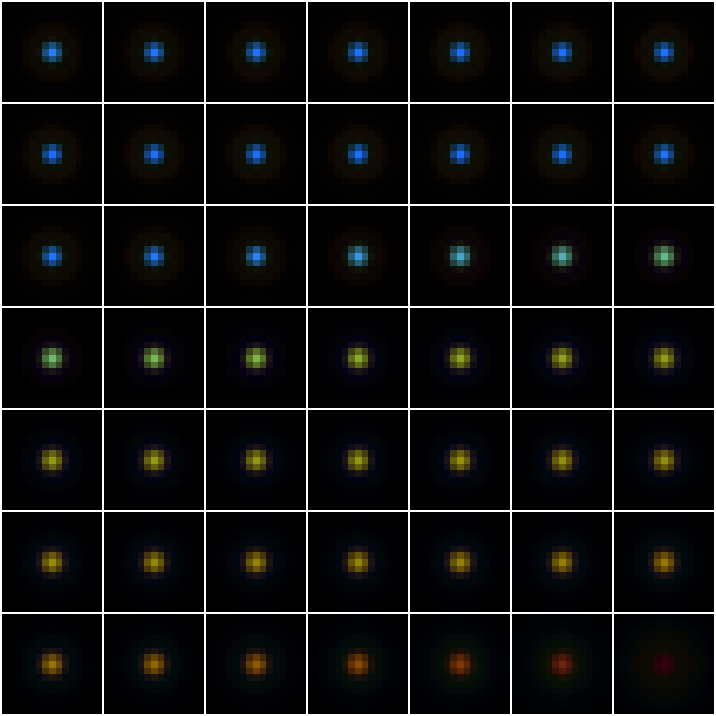}
   \caption{
   \op{
   \textbf{Learned PSF kernels.} 
   The time-variant camera PSF represented by 49 RGB color kernels, starts at the upper-left kernel (blue) and ends at the bottom-right kernel (red) in row-major order. These are a simulation of the camera PSF, computed as discussed in \cref{sec:PSFcomputation}}}
   \label{fig:grid_psf_kernels}
\end{center}
\end{figure}

\begin{figure}
\centering
    \includegraphics[width=1\columnwidth]{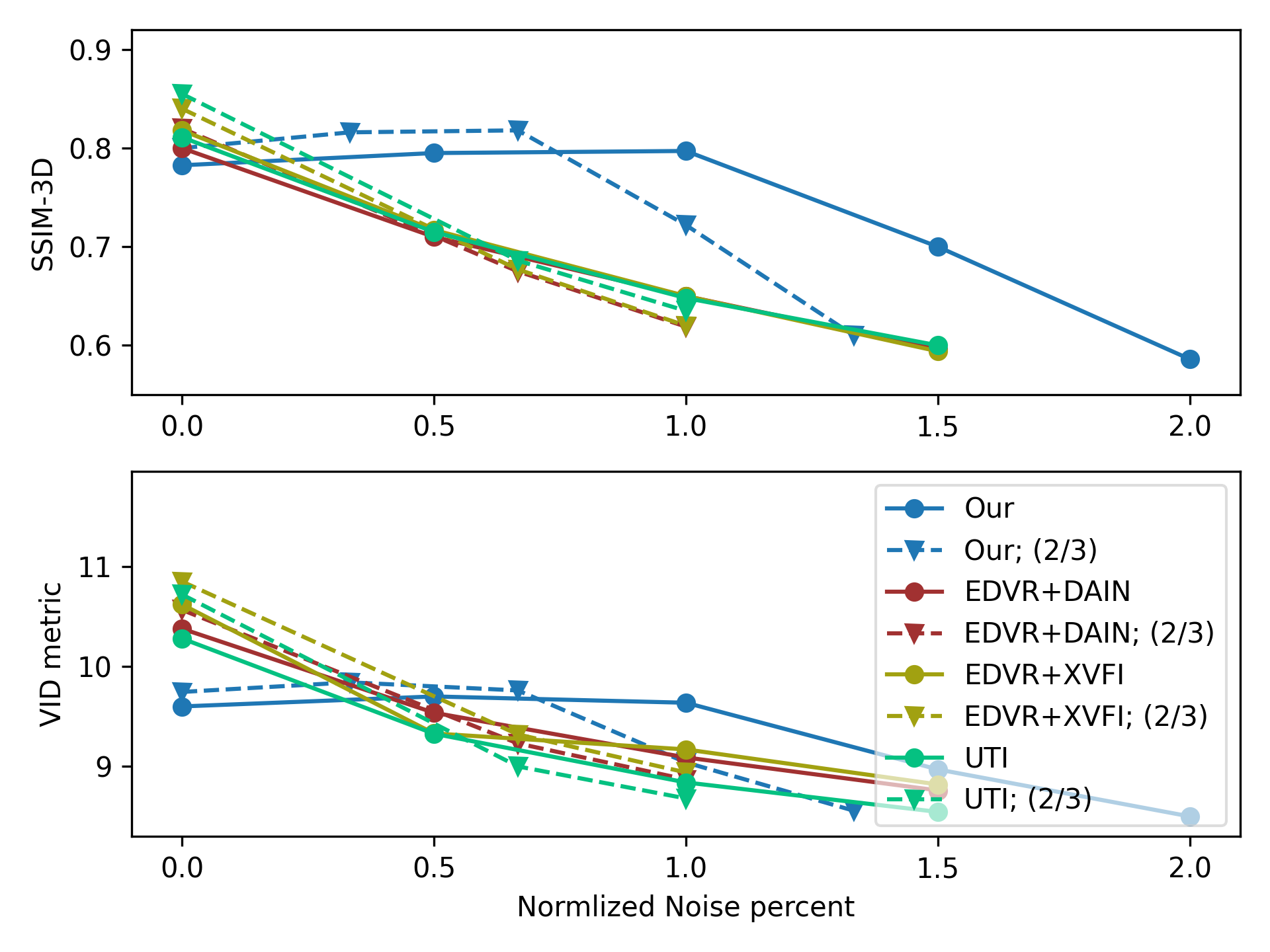}
\caption{\opb{\textbf{Video frame interpolation - SSIM3D and VID performance for different noise levels on REDS dataset \cite{Nah_2019_CVPR_Workshops}} Comparison of SSIM-3D metric and our VID metric for two exposure intervals: baseline and two-thirds of the baseline. The noise axis was normalized with respect to the exposure interval based on the SNR behavior (more information is described in \cref{SNRtradeoff}).
}}
\label{fig:vfi_ssim_reds}
\end{figure}

\section{Additional models architectures} \label{sec:additional_models}
\subsection{Deblurring models}
For the deblurring task, we train the same model architecture both for the linear code and the learned code (presented in \cref{miidleframe_table}). We used the proposed Unet model with a few minor adaptations due to the different task. Since the time parameter is irrelevant (always $t=0$) we did not concatenate a time channel to the input image. moreover, we replace AdaIN with group normalization since it is more stable during training, and set the groups number to 16.
\subsection{AdaGN instead of AdaIN video model}
In \Cref{table_ablation} we present a model with adaptive group normalization instead of AdaIN (config-g). In this architecture, each instance normalization was replaced using group normalization with 16 groups. Following the normalization we perform an affine transform, the same as performed in AdaIN. 
\subsection{Degraded architectures}
For the degraded models presented in \cref{fig:model_size_graph} we used Unet architecture \cite{Unet2015} in depth of 2 for the encoder and the decoder parts (two down/up-sampling), and with Mobile-Net convolution blocks \cite{mobilenet}. We added a skip connection from the input to the output, and neither added the time concatenated channel nor the positional encoding.
for B2D2 architecture we used double convolution blocks, while for B1D2 we used a single convolution block. each convolution consist of Inverted-Residual block \cite{mobilenet} followed by Leaky-Relu activation. the batch normalization in the Inverted-Residual block was replaced by AdaGN as described before. all the rest details remain the same as the proposed architecture.
}

\begin{figure*}
 \begin{center}
    \includegraphics[width=1\linewidth]{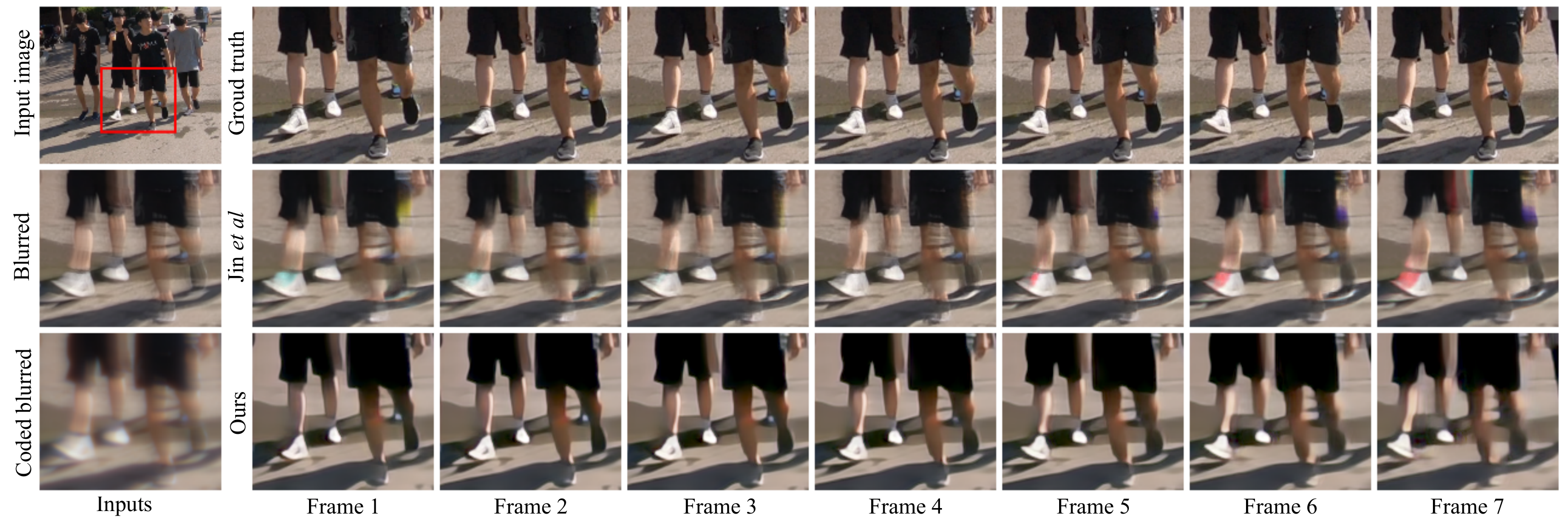}
   \caption{\textbf{Reconstruction performance (simulation) for seven frames.} (top row) GT image and zoom-in for a 7-frames burst, (middle row) conventional blur and  \cite{Jin_2018_CVPR} results, and (bottom row) our coded input and reconstruction results. The full result videos are presented in the supplementary video.}
   \label{fig:leg2}
\end{center}
\end{figure*}

\begin{figure*}
 \begin{center}
    \includegraphics[width=1\linewidth]{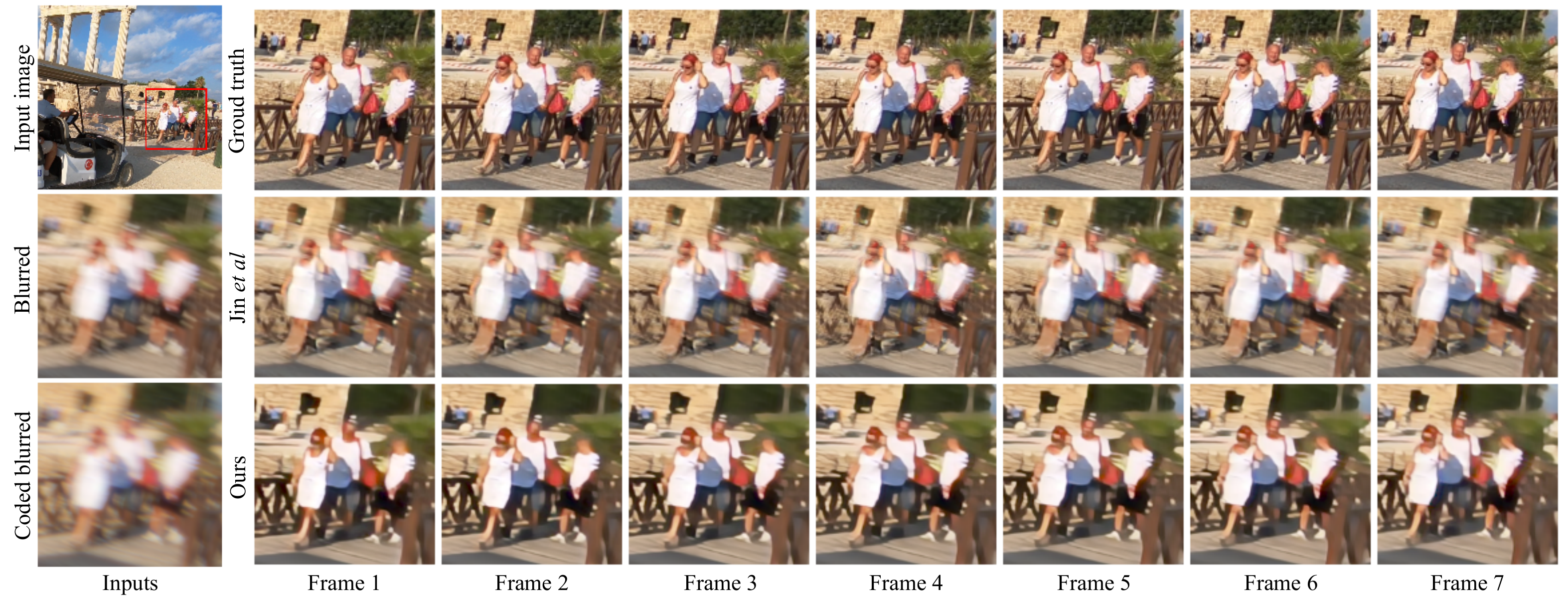}
   \caption{\textbf{Reconstruction performance (simulation) for seven frames.} (top row) GT image and zoom-in for a 7-frames burst, (middle row) conventional blur and  \cite{Jin_2018_CVPR} results, and (bottom row) our coded input and reconstruction results. The full result videos are presented in the supplementary video.}
   \label{fig:zabita}
\end{center}
\end{figure*}

\opb{
\section{Video Frame Interpolation Details} \label{sec:vfi_details}
For video frame interpolation model we used three consecutive coded blurred frames as an input to the model, and reconstruct a single frame in time interval $t\in[-1,1]$. The other details of training remain as in the image-to-video case. We set two video timing setups for the training and evaluation. Using 960 fps sharp video data set we generated "48-8" blurred video data by averaging 48 frames in linear space to a single blurred frame (equivalent to 50ms exposure), and passing 8 frames as reset time of the camera (intervals of 56 frames total). This timing setup is considered as our baseline. In the second timing setup, denoted as two-thirds of the baseline, we generate "32-16" dataset accordingly. The exposure interval is 33.3ms in this timing setup, which is two-thirds of the baseline. On training, we use batches of both timing setups to generalize the reconstruction for the different timings. Note that the $t$ parameter represents both intra- and inter-blur frame reconstruction for the specified interval, i.e. a sharp frame which is part of the blurred input image, and a sharp frame that is between blurred frames (the camera reset time).

\subsection{Adobe240 Dataset}
To verify our method on another dataset than the one we used for training and validation, we tested the models on the Adobe240 dataset \cite{su2017deep_adobe240} and presented our results in the main paper. Since this dataset is oriented for hand-held camera video deblurring, we picked 10 videos of dynamic scenes with less camera shake for our evaluation. Same as was done for REDS dataset, before creating the blurred frames by averaging consecutive frames we perform frame interpolation by a factor of 4 using \cite{DAIN} to get 960fps video. Inverse CRF was applied on the frame prior to the PSF convolution and temporal averaging (blurring) for blurred video simulation.

\subsection{Prototype Camera Video Acquisition}
We captured videos using our prototype camera with the temporal phase coding for each frame. We follow the baseline timing, namely 6/7 of the cycle is exposure time and 1/7 is for reset time. We capture in 4 fps, and the exposure time was set to $\sim$214ms accordingly. the reset time in this case is 35.7ms which is enough for the liquid lens to return to the initial state for the next frame capture. Even though high pfs was not our goal for the prototyping, the more prominent bottleneck in our case was the data transfer of the camera due to high-resolution frames acquisition (3200x2400) and not the liquid lens. As we did in the single image models, for the prototype reconstructions we trained a model with 3\% noise since real world images are noisier than the 1\% noise used for evaluation and comparisons. We placed colourful images on a rotating wheel to control the rotation speed. the images are under free use creative commons license from "pixnio" website. The captured blurred videos and the reconstructed videos are presented in the supplementary material.

\section{Results} \label{sec:Results}

In addition to the image-to-video results presented in the paper, we present additional  results in the supplementary video and here. The reconstructed videos were generated with 25 frames since we can choose any frames number using our time dependent CNN. The frame-rate difference compared to \cite{Jin_2018_CVPR} (which is limited to 7-frames only) is clearly noticeable in the supplementary video. A comparison between the frames in our results and \cite{Jin_2018_CVPR} are presented in \cref{fig:leg2}, \cref{fig:zabita} and \cref{fig:tiger}. Our method achieves improved results along the entire frame burst.
}

\begin{figure*}
\centering
\begin{tabular}{p{1cm}p{3cm}p{3cm}p{4cm}p{2cm}c}
  & Input & Our & EDVR+XVFI & UTI & EDVR+DAIN  \\
\multicolumn{6}{c}{
\includegraphics[width=1\linewidth]{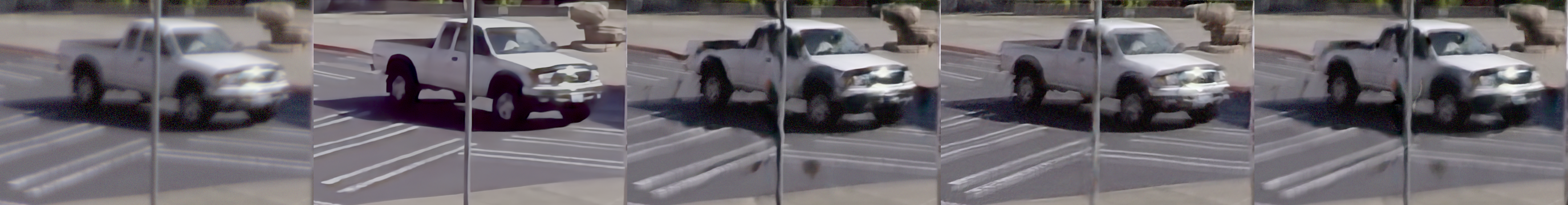}} \\
\end{tabular}
\caption{\textbf{Frame interpolation results.} For synthesized blurred input from Adobe240 dataset with noise $\sigma=0.5$ and the "48-8" timing setup. Left to right: Input blurred frame, our reconstruction sharp frame, and three alternative methods using a conventional camera.}
\label{fig:vfi_results_n05}
\end{figure*}

 \begin{figure*}
 \begin{center}
    \includegraphics[width=1\linewidth]{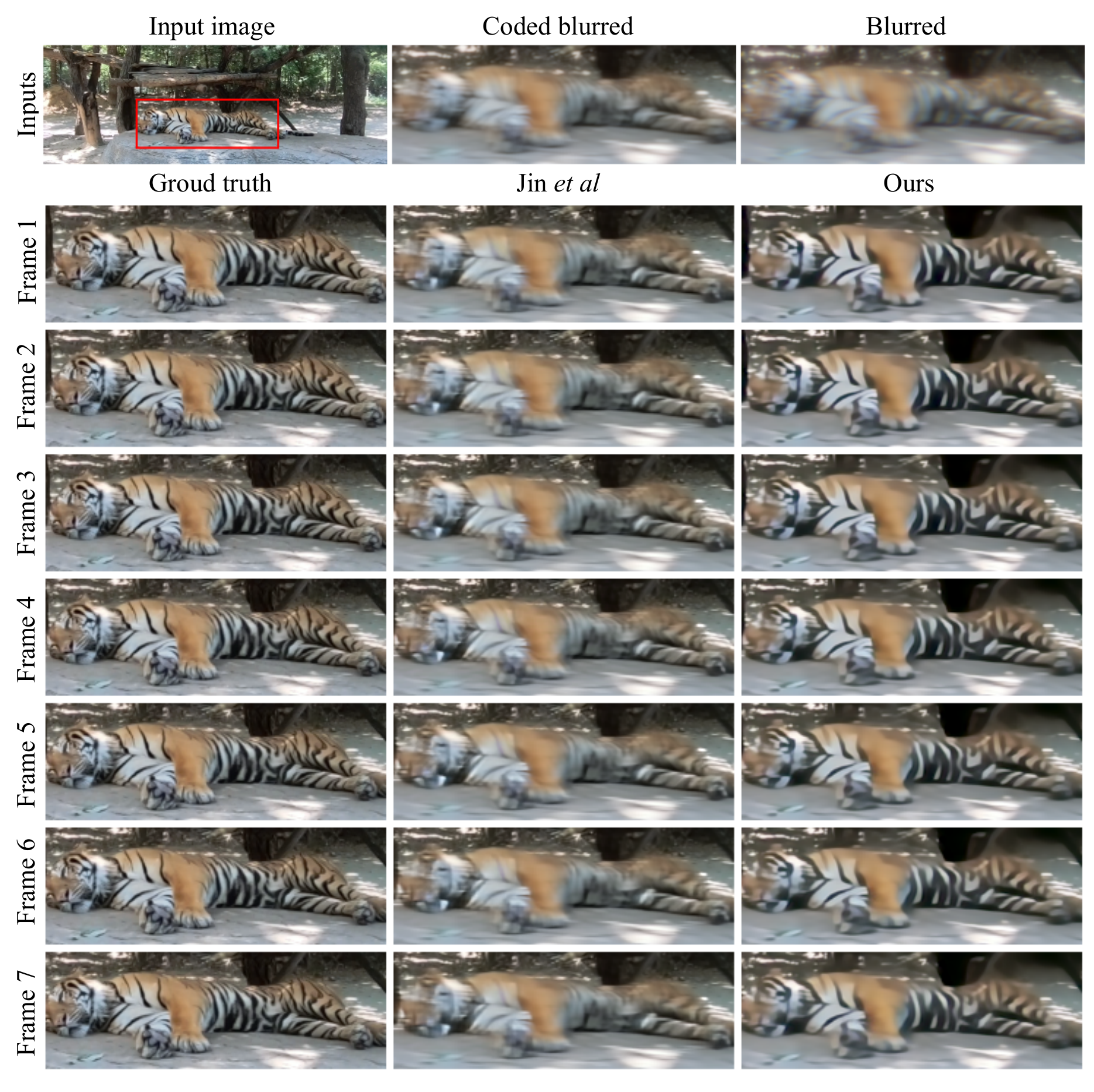}
   \caption{\textbf{Reconstruction performance (simulation) for seven frames.} (left column) GT image and zoom-in for a 7-frames burst, (middle column) conventional blur and  \cite{Jin_2018_CVPR} results, and (right column) our coded input and reconstruction results. The full result videos are presented in the supplementary video.}
   \label{fig:tiger}
\end{center}
\end{figure*}

\end{document}